 \documentclass[final, 5p, times, authoryear]{elsarticle}

\usepackage{longtable}
\usepackage{float}
\usepackage{adjustbox}
\usepackage{xltabular}
\usepackage{booktabs}
\usepackage{caption}
\usepackage{url}
\usepackage{graphicx}
\usepackage[colorlinks=true, allcolors=blue]{hyperref}
\usepackage{tcolorbox}
\tcbuselibrary{listings}

\usepackage{setspace}
\usepackage{amsmath}
\usepackage{cleveref}
\usepackage{amssymb}
\usepackage{multirow}
\usepackage[table,xcdraw]{xcolor}
\usepackage[normalem]{ulem}
\useunder{\uline}{\ul}{}
\usepackage{lineno}
\usepackage[authoryear, round]{natbib}
\usepackage[multiple]{footmisc}

\tcbuselibrary{breakable,listings,theorems}

\newtcbtheorem[number within=section]{mytheo}{Description}%
{colback=blue!5,colframe=blue!35!black,fonttitle=\bfseries}{th}

\usepackage{ragged2e} 

\newtcolorbox[auto counter, crefname={Prompt template}{Prompt templates}]{promptbox}{%
  colback=gray!5!white, colframe=black!40,
  title={Prompt Template 1},
  fonttitle=\bfseries, boxrule=0.5pt, arc=2pt,
  left=5pt, right=5pt, top=5pt, bottom=5pt,
  boxsep=3pt,
  breakable,
  before upper={\normalsize\setstretch{0.98}}%
}

\crefname{promptbox}{Prompt template}{Prompt templates}
\Crefname{promptbox}{Prompt template}{Prompt templates}

\crefformat{promptbox}{Prompt template~\ref{#2#1#3}}
\Crefformat{promptbox}{Prompt template~\ref{#2#1#3}}

\newtcolorbox[auto counter]{resultbox}[2][]{%
  colback=gray!5!white, colframe=black!40,
  title={Model Output~\thetcbcounter}, label={#1},
  fonttitle=\bfseries, boxrule=0.5pt, arc=2pt,
  left=5pt, right=5pt, top=5pt, bottom=5pt,
  boxsep=3pt,
  breakable,                 
  before upper={%
    \normalsize                  
    \setstretch{0.98}
  },
}

\newcommand{\softbox}[1]{%
  \begin{center}
    \setlength{\fboxsep}{8pt}
    \colorbox{gray!10}{%
      \begin{minipage}{0.8\linewidth}
        #1
      \end{minipage}
    }
  \end{center}
}
\newcommand{\q}[1]{``#1''}

\usepackage{glossaries}
\makeglossaries

\newacronym{LLM}{LLM}{Large Language Model}
\newacronym{SLM}{SLM}{Small Language Model}
\newacronym{CNN}{CNN}{Convolutional Neural Network}
\newacronym{NLP}{NLP}{Natural Language Processing}
\newacronym{ML}{ML}{Machine Learning}

\begin{document}

\begin{abstract}
Phishing websites pose a major cybersecurity threat, exploiting unsuspecting users and causing significant financial and organisational harm. Traditional machine learning approaches for phishing detection often require extensive feature engineering, continuous retraining, and costly infrastructure maintenance. At the same time, proprietary large language models (LLMs) have demonstrated strong performance in phishing-related classification tasks, but their operational costs and reliance on external providers limit their practical adoption in many business environments. 
This paper investigates the feasibility of small language models (SLMs) for detecting phishing websites using only their raw HTML code. A key advantage of these models is that they can be deployed on local infrastructure, providing organisations with greater control over data and operations.
We systematically evaluate 15 commonly used \Glspl{SLM}, ranging from 1 billion to 70 billion parameters, benchmarking their classification accuracy, computational requirements, and cost-efficiency. Our results highlight the trade-offs between detection performance and resource consumption, demonstrating that while \Glspl{SLM} underperform compared to state-of-the-art proprietary LLMs, they can still provide a viable and scalable alternative to external LLM services. By presenting a comparative analysis of costs and benefits, this work lays the foundation for future research on the adaptation, fine-tuning, and deployment of \Glspl{SLM} in phishing detection systems, aiming to balance security effectiveness and economic practicality.
\end{abstract}


\begin{frontmatter}


\title{Small Language Models for Phishing Website Detection: \\Cost, Performance, and Privacy Trade-Offs}




\author[1]{Georg Goldenits\corref{cor1}}

\cortext[cor1]{Corresponding author}
\ead{ggoldenits@sba-research.org}

\author[2]{Philip König}

\author[1]{Sebastian Raubitzek}

\author[1,2]{Andreas Ekelhart}



\affiliation[1]{organization = {SBA Research gGmbH}, 
                addressline = {Floragasse 7/5.OG},
                city = {Vienna},
                postcode = {1040},
                state = {Vienna},
                country = {Austria}}

\affiliation[2]{organization = {University of Vienna}, 
                addressline = {Universitätsring 1},
                city = {Vienna},
                postcode = {1010},
                state = {Vienna},
                country = {Austria}}





\end{frontmatter}




\section{Introduction}
Website phishing remains one of the most prevalent and damaging forms of online fraud, enabling attackers to steal user credentials, financial data, or other sensitive information by imitating legitimate websites~\cite{intro8,intro9}. The Anti-Phishing Working Group (APWG) reported over 1.1 million unique phishing websites detected in Q2 2025, representing a 13\% increase compared to Q1 in 2025, continuing a multi-year upward trend in attack frequency~\cite{intro1}. 
Unlike email phishing, web-based phishing exploits users' trust in domains, brands, and website structures, making it difficult to detect through superficial inspection~\cite{intro2}. As phishing websites become more sophisticated, evasion tactics become more prevalent and campaigns emerge rapidly, automated and adaptive detection techniques are essential~\cite{intro9,JNCA2}.

Classical machine learning and deep learning approaches for website phishing detection, such as URL-based classifiers~\cite{intro5,intro6} or visual similarity models based on \Glspl{CNN}~\cite{intro3,intro4,Phishpedia} have shown promising detection rates under controlled conditions. However, they face several limitations in practice. Models trained on static features are vulnerable to adversarial evasion, such as obfuscating malicious scripts or dynamically generating HTML content~\cite{intro11,intro12,intro10}. Furthermore, \Gls{ML} classifiers require continuous retraining or feature updates to remain effective against rapidly evolving phishing tactics, leading to high operational costs. Challenges such as high false-positive rates, susceptibility to adversarial attacks, scalability issues, and data imbalances limit the usefulness of \Gls{ML}-based detection approaches~\cite{intro7,JNCA3}. These factors underscore the need for more adaptable and robust detection mechanisms.

Recent work has proposed using \Glspl{LLM} for phishing website detection by treating the HTML structure and embedded text as natural language input. Proprietary \Glspl{LLM}, such as GPT-4 or Gemini, have demonstrated strong performance in classifying phishing websites and even generating natural-language explanations of their reasoning \cite{PhishLLM,propBM1}. However, these approaches are often impractical in commercial settings, as they depend on vendor-controlled APIs, incur high inference costs, and make fine-tuning or customisation expensive. They also pose data privacy risks, as sensitive information must be sent to external systems, and may suffer from availability issues due to varying request latencies that can range from few seconds to minutes~\cite{intro14, intro13}.

\Glspl{SLM}, which we define as models with $\leq$70b parameters, present a promising alternative for phishing website detection, especially when they are open source and deployed locally. They offer several advantages, including significantly reduced operating costs, independence from vendor APIs, improved data privacy and confidentiality, easier compliance, and the ability to fine-tune models on-premise for domain-specific needs. However, to the best of our knowledge, no study outlines the utility, performance, costs, and benefits of using \Glspl{SLM} in comparison to large proprietary models.

In this study, we systematically benchmark a diverse set of 15 commonly used \Glspl{SLM}, ranging from 1 billion to approximately 70 billion parameters, for website phishing detection. The selected models include different sizes of Gemma\footnote{\url{https://ollama.com/library/gemma3}}, Deepseek\footnote{\url{https://ollama.com/library/deepseek-r1}}, Qwen\footnote{\url{https://ollama.com/library/qwen3}} and Llama\footnote{\url{https://ollama.com/library/llama3.1}, \url{https://ollama.com/library/llama3.2}, \url{https://ollama.com/library/llama3.3}}, as well as models that are only available in one size, such as Mistral\footnote{\url{https://ollama.com/library/mistral-nemo}}, gpt-oss\footnote{\url{https://ollama.com/library/gpt-oss}}, phi\footnote{\url{https://ollama.com/library/phi3}}, and dolphin\footnote{\url{https://ollama.com/library/dolphin3}}. 
We evaluate those models along three axes: time to analyse website HTML and produce results, phishing classification correctness, and output coherence. 
To generate this benchmark, each model classifies 1,000 websites (500 phishing and 500 benign) sampled from a freely available dataset~\cite{data}.

To determine the suitability of local \Glspl{SLM} for phishing detection, we structure our approach along the following research questions:

\begin{itemize}
    \item \textbf{RQ1}: Do \Glspl{SLM} deliver coherent and reliable outputs, including meaningful reasoning for phishing website detection?
    \item \textbf{RQ2}: How do the costs and benefits of local models compare to state-of-the-art proprietary models such as GPT4?
    \item \textbf{RQ3}: How does the length of the HTML code influence both analysis runtime and detection performance?
\end{itemize}

The contributions of this article are:
    \begin{itemize}
     \item We benchmark 15 \Glspl{SLM}, regarding various numeric and text-based criteria on the task of website phishing detection to develop a detection method that can compete with state-of-the-art larger proprietary \Glspl{LLM}.
     \item The benchmarking methodology, dataset, and source code are publicly available on Github\footnote{\url{https://github.com/sbaresearch/benchmarking-SLMs}}, so other researchers and practitioners can reproduce the results and test new models in the future.
     \item We discuss benefits and challenges of small local models compared to commercially available options and provide recommendations for application of \Glspl{SLM} for website phishing detection.
 \end{itemize}

This work is structured as follows: \Cref{sec:relW} presents related work focusing on using \Glspl{LLM} for website phishing detection. In Section 3, the experimental design setup is presented, with Section 4 containing the results of the experiments. Section 5 discusses the advantages and costs of local \Glspl{LLM} compared to commercial applications, providing a foundation for application recommendations. Section 6 concludes the work and outlines future research directions.

\section{Related Work}\label{sec:relW}
Although the field of \Gls{LLM}-based website phishing detection has only recently gained significant traction, several studies have reported promising results using different models and analysis strategies. Table~\hyperref[tab:lit_overview]{\ref{tab:lit_overview}} provides an overview of these studies,  highlighting input data modalities, models, and their reported classification performance.

\begin{table}[ht]
\centering
\caption{Summary of related work classified into data modes, \Glspl{LLM} used, dataset size, and reported classification performance (F1-score).}
\label{tab:lit_overview}
\resizebox{\columnwidth}{!}{%
\begin{tabular}{@{}lllll@{}}
\toprule
\textbf{Publication} &
  \textbf{Input Data} &
  \textbf{\Gls{LLM}} &
  \textbf{Dataset Size} &
  \textbf{F1-score} \\ \midrule
\cite{propBM} &
  Text + Vision &
  \begin{tabular}[t]{@{}l@{}}Gemini Pro-Vision 1,\\ GPT-4-turbo,\\ Claude 3 Opus\end{tabular} &
  \begin{tabular}[t]{@{}l@{}}$\sim$3,000 benign\\ $\sim$1,500 phish\end{tabular} &
  0.81 - 0.92 \\
\cite{propBM1} &
  Text + Vision &
  \begin{tabular}[t]{@{}l@{}}GPT-4;4V;4o,\\ GPT-3.5-turbo,\\ Gemini Pro 1.0,\\ Command R+,\\ Llama 2 70b,\\ Llama 3 70b,\\ Gemma 2 9b\end{tabular} &
  \begin{tabular}[t]{@{}l@{}}1,000 benign\\ 1,000 phish\end{tabular} &
  0.72 - 0.99 \\
\cite{RLWork2} &
  Text &
  \begin{tabular}[t]{@{}l@{}}Qwen2.5-vl.72b-instruct,\\ Gemini-2.0-Flash,\\ GPT-4o,\\ GPT-4o-mini\end{tabular} &
  \begin{tabular}[t]{@{}l@{}}500 benign\\ 500 phish\end{tabular} &
  0.85-0.97 \\
\cite{RLWork3} &
  Text &
  \begin{tabular}[t]{@{}l@{}}GPT-3.5-turbo\\ Claude 2\end{tabular} &
  \begin{tabular}[t]{@{}l@{}}500 benign\\ 500 phish\end{tabular} &
  0.78-0.93 \\
\cite{RLWork4} &
  Text + Vision &
  Gemini-1.5-Flash &
  \begin{tabular}[t]{@{}l@{}}17,794 benign\\ 9105 phish\end{tabular} &
  0.75-0.98 \\
\cite{PhishLLM} &
  Text + Vision &
  GPT-3.5-turbo-16k &
  \begin{tabular}[t]{@{}l@{}}6,000 benign\\ 6,000 phish\end{tabular} &
  0.86

\end{tabular}
}
\end{table}

Regarding input data, two main approaches are commonly used in the literature: text-only, which consists of website HTML code and URLs, and text combined with screenshot images. The choice of data modality typically depends on data availability and the capabilities of the selected models. Since most \Glspl{LLM} are designed to process and generate text, relying solely on HTML code is frequently chosen as a starting point for a phishing detection framework. Combining text and images is a common next step and a method often used to detect brand impersonations.

The most frequently used models are versions of GPT-3.5 and GPT-4, followed by iterations of the Gemini models. Open-source alternatives, such as Llama and Qwen, are less commonly chosen.
The dataset sizes used in the different studies vary but generally range of a few thousand websites. Typically, researchers use balanced datasets, with an equal number of phishing and benign websites.
As a comparative performance measure in phishing detection setups, the F1-score is commonly reported. 
The reported F1-scores range from 0.72 to 0.99.
Detailed descriptions of the cited detection architectures can be found below.

\cite{propBM} employ a multimodal approach that involves both brand identification and domain verification. In the first step, \Glspl{LLM} must identify the brand based on the website's HTML code and a screenshot of the website. The recognised brand, combined with the input domain name, then serves as input for the second \Gls{LLM}, which is tasked with classification.
Regarding the prompt design, there are three different versions depending on the input data mode (Screenshot, HTML, Screenshot + HTML), which instruct the \Gls{LLM} to analyse the given input and return a list of indicators, like \texttt{brand}, \texttt{has credentials} or a \texttt{confidence score}.
The dataset used in their work is freely available in a GitHub repository\footnote{\url{https://github.com/JehLeeKR/Multimodal_LLM_Phishing_Detection/tree/main/data/MMLLM_Benign/1}}

\cite{propBM1} use a different approach for handling the website screenshot, utilising an OCR model to extract text elements from the image. The extracted elements, along with the HTML code, are simplified and combined with the full screenshot and the URL, forming a data foundation for the analysis.
The prompt instructs the \Gls{LLM} to analyse the input data and gives a detailed step-by-step description of how to proceed. The output should then be structured as a JSON-style document containing certain keys, such as 'phishing score', 'brands', or 'suspicious domain'.
Although the method for deriving the dataset is provided, the final dataset is not freely available.

Compared to other approaches that rely on a single agent (or multiple non-collaborating agents), \cite{RLWork2} present a multi-agent system where each model represents a different expert to analyse website HTML code, content, URL, and screenshot. A moderator agent tracks if a common answer has been found, and if yes, provides the result to a judge agent who makes the final decision.
The prompt template is minimalistic compared to other approaches, but also instructs the model to give a structured list of items as a response.
The dataset is derived from various sources that can be accessed for free, but the final version is not freely available.

Unlike other related work, \cite{RLWork3} do not build an involved framework to detect malicious websites, where each website goes through in-depth analysis steps involving URL, HTML code, brand identification, or screenshot. Instead, in their approach, 50 URLs are batched together and are simultaneously analysed by an \Gls{LLM}. To enhance model performance, a classification head is added to the model architecture, whose weights are trained using a sample of the utilised URL dataset.
The base dataset can be found in an online repository \cite{RLWork3.1}, and while the exact sample used in their work is not directly available, the sampling was performed using a random seed, allowing for reproducibility.
Three versions of the prompt were tested: a zero-shot version, a role-playing version, and a chain-of-thought version, to assess the output quality for different prompt styles. All prompts instruct the model to output only a Boolean decision, indicating whether a URL corresponds to a phishing website or not.

\cite{RLWork4} have a different focus, testing the resilience of phishing detection systems to evasion strategies.
Based on the original HTML code of legitimate websites, deviating versions are created, which incorporate phishing techniques. The dataset containing these changed websites is available in a GitHub repository~\footnote{\nolinkurl{https://github.com/LetsBeSecure/PhishOracle-Project?tab=readme-ov-file}}. 
The detection performance of the models is assessed based on all the available website versions. All the tested approaches are prone to evasion techniques, thus losing classification performance; however, \Glspl{LLM} show the strongest resilience to HTML perturbations and are deemed the most resilient of the tested techniques. However, the performance loss of the \Gls{LLM} still shows the need to further improve upon the existing detection techniques.

A different direction is taken by \cite{PhishLLM}, who emphasise brand recognition as a mechanism to collect potential URLs a brand might use and also account for changes such as logo or web presence that a brand might have gone through. This is done to build a dependable database for each brand, with the goal of reducing detection errors and minimising reliance on a reference database. 
In contrast to other works, which only use website screenshots, a system is built to iteratively navigate to credential-taking pages when starting on a non-credential-taking page. 
In their pipeline, \Glspl{LLM} are used to detect the brand and predict whether a website is credential taking or not. For both tasks, the prompting technique relies on structured prompts including few-shot examples.
The datasets used to answer their research questions can be freely downloaded from a repository \footnote{\url{https://sites.google.com/view/phishllm/experimental-setup-datasets}}.

While this work focuses solely on detecting suspicious websites, adversarial phishing attack vectors also include emails and SMS messages. While \Glspl{LLM} get used for both website analysis and email/SMS analysis, the detection goals differ \cite{RLWork1,RLWork5,RLWork6}. With emails and SMS, the focus is on the used language, potential URLs in the text, and the sender.
In contrast, website phishing detection goes beyond just human language analysis, as HTML code and structure hold crucial information for detecting malicious websites.

\section{Methodology}
\begin{figure}[ht!]
    \centering
    \includegraphics[width=1\linewidth]{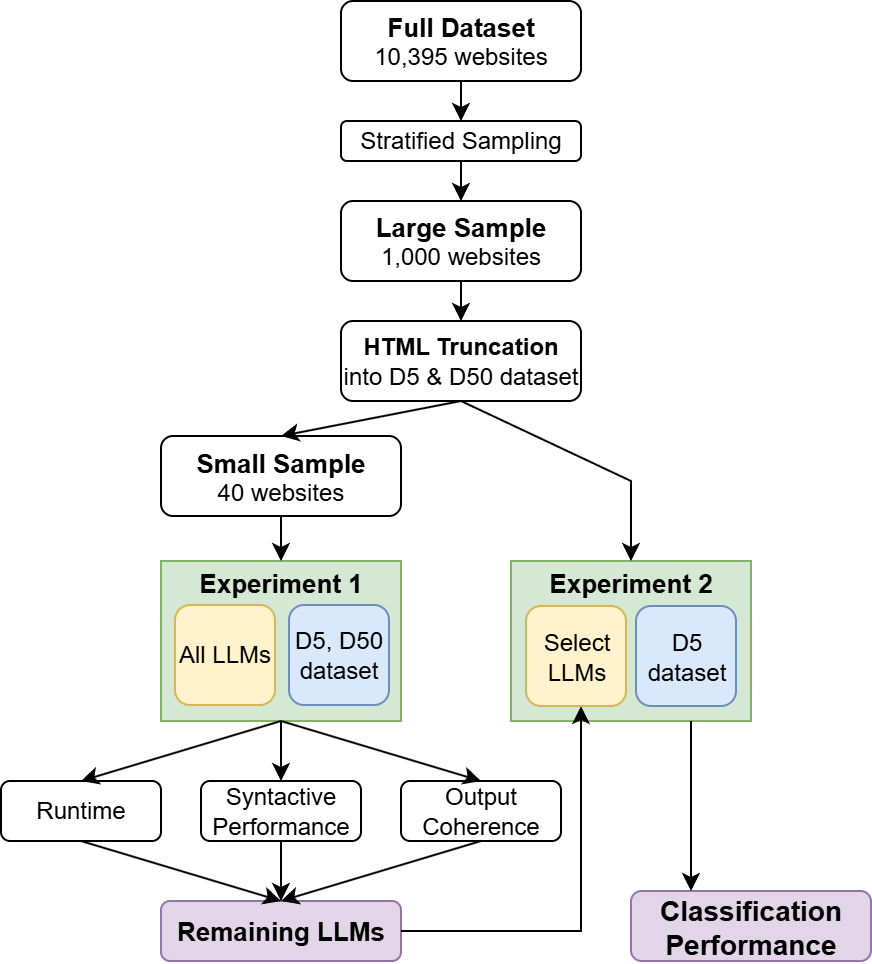}
    \caption{Methodological approach of the \Gls{LLM} benchmarking pipeline.}
    \label{fig:MethodStages}
\end{figure}

This work examines the costs and benefits of using small local \Glspl{LLM} for phishing detection compared to commonly used proprietary models. To gain detailed insights into the advantages and disadvantages of using such models and comparing them to proprietary models, two experiments are conducted, which cover different input data and data processing steps. In the Model Selection experiment (Experiment 1), the behaviour of a local \Glspl{LLM} is examined on a small subset of the full dataset used in the Scaled Evaluation experiment (Experiment 2). The goals of the first experiment are to uncover random behaviour, model runtime and model output quality in terms of completeness and formatting. In experiment 2, models that performed satisfactorily in experiment 1 are used to classify a larger set of websites. In this experiment, the focus lies on the actual phishing detection performance, output coherence and the incurred analysis cost.
{\Cref{fig:MethodStages}} summarises the experiments, dataset and models that are subsequently described in detail.

\subsection{Dataset}
The dataset used as a basis for this work can be accessed freely online and is provided to improve phishing detection techniques \cite{data}. The full dataset contains detailed information on 10,395 websites, comprising 5,244 legitimate and 5,151 phishing websites. This dataset includes HTML files, URLs, screenshots, CSS files, and information about JavaScript components and their execution. In this work, only the index HTML file, which displays the page with all JavaScript code executed, was selected, while the rest of the data was discarded. Focusing only on the HTML file ensures that our approach is comparable to publications mentioned in the related work and reproducible, as scraping HTML code is easier than accessing all the files in the background of the website.

The data repository was chosen because, with approximately 7,000 downloads, it is frequently used and contains an extensive data collection.

Based on this initial dataset, a stratified sample of 1,000 websites is selected, which is then further processed for use in the two experiments. The processing steps are described below.
Data reduction is necessary to ensure a feasible workload in terms of available computational resources and the resulting analysis runtime.

\begin{itemize}
    \item \textbf{Experiment 1}: For the initial Model Selection experiment, we sampled 20 benign and 20 phishing websites of the 1,000 website large sample using a stratified random sampling approach. This ensured that the dataset covers a broad range of HTML document lengths, thereby reducing the risk of bias in the evaluation. For the sampled websites, we created two different truncated versions, referred to as the D5 and D50 datasets. Both datasets contain only the HTML code elements most relevant for phishing detection, as these provide key information about the contents and structure of the websites \cite{propBM1}. The two datasets differ in the amount of HTML retained: the D5 dataset includes only a maximum of 5\% of the original websites' HTML code, while the D50 dataset includes a maximum of 50\%. 
    This truncation serves two purposes: 1) to avoid information that is unnecessary for phishing detection, 
    and 2) to reduce the computational costs when processing the data with \Glspl{LLM}.
    
    To identify the most relevant HTML code elements, we built on the work of \cite{propBM1}, who selected the following tags for a similar phishing detection task: \textit{head, title, meta, body, h1, h2, h3, h4, h5, h6, p, strong, a, img, hr, table, tbody, tr, th, td, ol, ul, li, ruby, and label}. Our truncation approach extends this by 
    ranking the tags according to their importance for phishing detection \cite{RLWork4}. Specifically, the elements: a, img, meta and link are deemed the most important, followed by HTML and head, while all remaining elements were assigned the lowest importance. Typically, these elements convey the most expressive information about a website, including brand cues, indications of potential phishing links, and high semantic meaning. The truncation process works as follows: first, it unwraps all elements, and then removes empty elements. In an iterative reduction, the least important remaining elements are removed until the HTML token count falls below the target threshold, either 5\% or 50\% of the total token count. If the removal of a component would eliminate an entire class of elements, elements from other classes with multiple occurrences are removed first. The only difference between the D5 and D50 dataset truncation processes is that the allowed HTML token threshold is higher for the D50 dataset.
    
    \item \textbf{Experiment 2}: In a setup similar to Experiment 1, a larger dataset consisting of 500 benign and 500 phishing websites was selected using a stratified random sampling strategy to ensure a balanced and representative distribution across different website characteristics. Based on the findings from Experiment 1, this dataset was truncated following the D5 dataset, retaining a maximum of 5\% of the original HTML content per website.
\end{itemize}

It is noteworthy that balanced datasets were used in both experiments, even though in real-world scenarios, the vast majority of websites encountered are expected to be benign. Unlike supervised learning approaches, \Glspl{LLM} analyse each website independently and do not rely on aggregated training data. Consequently, when websites are analysed in isolation rather than within a continuous conversation context, there is no need to reflect the real-world class distribution. Using a balanced dataset for analysis in this setting offers two key advantages: 1) it simplifies the interpretation of classification metrics, as results are not affected by class skew, and 2) it ensures a sufficient number of phishing samples, enabling more robust analysis of the \Glspl{LLM}' phishing detection capabilities.

\subsection{Model Selection}
Given the large number of local \Glspl{LLM}---Ollama alone lists over 1,000 models, versions, and specialisations---it is infeasible to evaluate all options exhaustively~\footnote{https://ollama.com}. Therefore, a subset of models was selected for this study, based on clearly-defined inclusion and exclusion criteria, summarised in \Cref{tab:llmSelect}.

\begin{table}
\centering
\caption{\Gls{SLM} selection criteria}
\label{tab:llmSelect}
\footnotesize
\begin{tabular}{p{\linewidth}}
\toprule
\multicolumn{1}{c}{\textbf{\Gls{SLM} Selection Criteria}}                  \\ \midrule
In the top 30 \q{Popular} models on Ollama in July 2025                \\ \midrule
Model has approximately 70 billion parameters or fewer \\ \midrule
Model has 128,000 context window length \\ \midrule
If there are multiple model variants available, choose the smallest model, a model in the centre of the parameter range and the model closest to 70b
\\ \midrule
Exclusion if tagged as \q{embedding} \\ \midrule
If the exact popular model does not have the required context window length, choose a model of the same family that falls within the scope (e.g., mistral:7b's context window is too short, therefore, it is replaced by mistral-nemo) \\ \bottomrule
\end{tabular}
\end{table}

Two key constraints informed our selection process. First, only models up to 70 billion parameters were included to focus on \Glspl{LLM} that can run locally on limited hardware. This allows organisations to retain sensitive data on-premise, thereby addressing privacy and compliance concerns, while also reducing operational costs. Furthermore, studies evaluating commercial \Glspl{LLM}, such as GPT models, for phishing detection already exist (see~\Cref{sec:relW}). Hence, our work focuses on local, open-source alternatives to provide a complementary perspective.
Second, a minimum context window of 128,000 tokens was chosen to ensure that the HTML input could be fully processed by the \Glspl{LLM}, avoiding automatic truncation by the model itself. Despite the truncation process, the HTML code often remains very long.
Allowing models with smaller windows to internally truncate prompts would compromise comparability.
Based on the above model selection criteria, 15 models were selected, spanning a diverse range of architectures and parameter sizes:
deepseek-r1: 1.5b, 14b, 70b; gemma3: 4b, 12b, 27b; qwen3: 4b, 30b; llama: 3.2:1b, 3.1:8b, 3.3:70b; mistral-nemo:12b; phi3-medium: 14b; dolphin3: 8b and gpt-oss: 20b.\\
\textbf{Considerations:}
\begin{itemize}
    \item Since no singular llama3.x model on Ollama comes with all required parameter sizes; multiple versions were included to represent a range of models. 
    \item The chosen Qwen3 models feature an exceptionally large context window length of 256,000 tokens, which allows them to process more input, compared to the other selected models. To maintain comparability, the maximum prompt length was capped at 128,000 tokens, ensuring that no model receives additional information.
    \item Since \Gls{LLM}-based phishing detection research often relies on GPT-family models, we decided to include gpt-oss:20b as an open-source GPT alternative with a comparable model size to the other models tested in our experiments.
\end{itemize}

\subsection{Prompt Design}\label{sec:prompt}
Designing an effective prompt is a critical step in \Gls{LLM}-based phishing website detection. The primary objective of the prompt is to instruct the model to analyse a given HTML document and corresponding URL, identify potential phishing indicators, and produce a structured, interpretable output that can be used for quantitative evaluation. Using HTML and URL is common practice for LLM-based phishing detection but also other approaches \cite{JNCA1}.
The final prompt, shown in Prompt Template 1, is based on the template proposed by \cite{propBM1} and was iteratively refined through multiple design cycles.
The prompt consists of three core elements: 1) the task description, which instructs the model to classifying a website as benign or phishing based on the provided HTML code and URL, 2) guidance examples (few-shot prompting) that outline common phishing indicators to guide the model's reasoning, and 3) the output specification, which defines the required JSON output format including fields and score ranges. While the models are clearly instructed on how to deliver their output, any JSON-parsable response is accepted during the analysis phase. We anticipate inaccuracies when working with smaller models, and therefore do not discard results solely due to minor formatting deviations.  

The following considerations shaped the final design:
\begin{itemize}
    \item \textbf{Flat Prompt Design}: 
    We use the generate mode (includes no conversational history), since each website must be analysed independently.
    As a result, the prompt had to be written in a flattened format, i.e., without explicit role labels like system or user.

    \item \textbf{Few-Shot Prompting}: 
    To guide the model in detecting phishing characteristics, the prompt includes a limited number of examples of common phishing indicators (e.g., suspicious URLs, hidden forms, misleading redirects).
    These examples provide context and decision-making cues.

    \item \textbf{Output Formatting}: 
    A strict JSON-parsable output format was defined to standardise model responses. The output contains three fields: \texttt{phishing\_score} (integer from 0 to 10, segmented into predefined risk brackets to reduce ambiguity and facilitate consistent interpretation across different models). \texttt{is\_phishing} (boolean classification), \texttt{reasoning} (brief textual explanation citing key indicators).\\
    Furthermore, the output length is limited to 1,000 tokens in an attempt to encourage models to retain the required output formatting and obtain concise answers.

\end{itemize}


\begin{promptbox}\label{pr:one}
    You are a cybersecurity expert analysing websites for phishing attempts. Your task is to examine the provided HTML code and the original website URL to determine if the website is likely a phishing site.\\
\newline
    **Important:** The HTML may be truncated to reduce costs, so CSS styles and JavaScript code may be missing. Focus on the HTML structure, text content, URLs, and the original URL.\\
    Original URL: '\{original\_url\}'\\
    HTML: '\{html\_text\}'\\
    Original HTML character count: '\{original\_character\_count\}'\\
\newline
    When assessing the likelihood of phishing, consider both your general knowledge of phishing techniques and the evidence you see in the data - focus on HTML structure, content, and domain/URL consistency.\\" 
    Below are common examples of phishing indicators you may find useful, but you are not limited to them:\\
    \newline
    1. **Suspicious URLs/domains**\\    
    2. **Form analysis**\\   
    3. **Content and language**\\   
    4. **HTML structure issues**\\  
    5. **Link analysis**\\  
    6. **Brand impersonation**\\
    7. **Missing legitimacy markers**\\
    
    \textit{\textcolor{red}{Indicator details have been omitted for brevity. See~\ref{appendix} for a full version.}}\\
    
    **Note:** Since CSS/JS may be truncated, focus on HTML tags, text content, and URL analysis rather than visual styling or dynamic behaviour.\\
    **Scoring guide:**\\
    - 0-2: Very unlikely phishing (legitimate site)\\
    - 3-4: Low risk (minor suspicious elements)\\
    - 5-6: Medium risk (several concerning indicators)\\
    - 7-8: High risk (multiple clear phishing signs)\\
    - 9-10: Very high risk (obvious phishing attempt)\\
   \newline     
    **Required output format (JSON only):**\\
    \{\\
        phishing\_score: int [0-10],\\
        is\_phishing: boolean [true/false],\\
        reasoning: string [Brief explanation of your decision based on specific indicators found]\\
    \}\\
    \newline    
    **Output Constraints:**\\
    Do only output the JSON-formatted output and nothing else.

    \end{promptbox}

\subsection{Used Hardware}
For all computations, initially, the Nvidia A100 80GB PCIe GPU and an AMD EPYC 9554 CPU were tested. For the 70b models, a switch to the Nvidia H100 80GB PCIe GPU was necessary. All experiments were conducted on cloud infrastructure from runpod.io\footnote{\url{https://www.runpod.io/}}. 

\subsection{Experimental Design Setup}
The goal of this work is to evaluate the performance of \Glspl{LLM} for phishing detection by assessing their output quality and correctness, runtime, and variability. To ensure a clear structure for this work, the evaluation process is divided into two experiments. The first experiment focuses on randomness, runtime, and output completeness, establishing a baseline for model stability and feasibility. The second experiment then evaluates the models' phishing classification performance and computational costs.

\subsubsection{Experiment 1 - Syntactic Performance and Runtime}\label{Exp1Method}
When using \Glspl{LLM}, randomness is critical factor, especially when aiming to build a reliable and reproducible. Output variability arises because models predict the next output token based on a probability distribution learned during training. Even when presented with identical prompts, the model may select different tokens, leading to variations in output~\cite{googleLLMIntro,transformer}.

To mitigate this behaviour, most \Glspl{LLM} have specific parameters, such as \texttt{temperature}, \texttt{top\_p}, and \texttt{top\_k}, that can be adjusted to reduce output variance.
\softbox{
  \textbf{Temperature:} Values near 0 make outputs deterministic by always selecting the most probable token, while values near 1 increase variability by allowing lower-probability tokens to be chosen.

  \textbf{Top\_p:} Sets a cumulative probability threshold. The model samples only from the smallest set of tokens whose combined probability reaches this threshold.

  \textbf{Top\_k:} Limits selection to the k most probable tokens. A value of 5 means that only the top five tokens are considered.
}
\noindent
For our experiments, the parameters for each model are set to temperature = 0, top\_p = 0, top\_k = 1, ensuring outputs are as deterministic as possible.

Using these settings, each model uses the prompt from Section \Cref{sec:prompt} and analyses both the D5 and D50 phishing websites datasets, truncated to different levels (5\% and 50\% of the original HTML length).
Each model analyses the same website five times to measure model output variability and to estimate the total runtime. This results in 3,000 total runs per dataset (40 websites x 15 models x 5 repetitions). For testing the variability and stability of the model results, using the small sample dataset already provides valuable insights that can be applied to larger samples without requiring days of waiting for the results to be computed. To illustrate the point: If the full 1,000 website sample were chosen, instead of 3,000 analysis runs, this number would drastically increase to 75,000. 

The goals for this experiment are to answer the following questions:

\begin{itemize}
    \item Is the output deterministic or do variations occur?
    \item How long are the model runtimes, and how does the prompt length influence it?
    \item To what extent do models adhere to the specified JSON output format?
\end{itemize}

At the end of Experiment 1, models that fail to produce consistent and correctly formatted output or cannot run efficiently on the available hardware are excluded from Experiment 2.

\subsubsection{Experiment 2 - Phishing Detection Performance}
The findings of Experiment 1 directly inform the design of Experiment 2. Only the models that passed Experiment 1 are included in the analysis, ensuring the evaluation focuses on practical and reliable candidates.
At this stage, the full data sample of 1,000 websites is utilised, comprising 500 benign and 500 phishing sites. Due to computational resource constraints, only the D5 version (truncated to 5\% of the original HTML length) is utilised. The prompt remains identical to the one from Experiment 1 (see \Cref{sec:prompt}).

The primary goal of this experiment is to evaluate the classification performance of the models using standard metrics such as accuracy, precision, recall, and F1 score. Furthermore, the output coherence is examined by comparing the output variables to ensure that they lead to consistent decisions. For instance, if the boolean phishing variable is \texttt{False}, a corresponding low phishing score should be obtained, and the reasoning should reflect this decision.
Finally, the experiment assesses the computational costs of running these models, allowing a comparison to commercial approaches. Ultimately, the experiment provides insights into the practical trade-offs between small open-source \Glspl{LLM} in phishing detection and commercial alternatives.


\section{Results}
This section presents the results of the two experiments. The evaluation experiment (Experiment 1) serves as the setup and feasibility study, focusing on runtime performance, output correctness in terms of the required formatting, and output coherence relating to the required output parameters, which means all parameters should indicate the same decision regarding the website classification. A discussion of the implications of the results for the setup of Experiment 2 follows this. 
Experiment 2 evaluates the classification performance of the selected \Glspl{LLM} on the phishing detection task.

\subsection{Experiment 1}
The goal of the first experiment is to assess the practicality of different \Glspl{LLM} for phishing detection scenarios. 
To this end, the focus lies on three key aspects: runtime performance, output variability, and output correctness. These factors offer initial insights into the stability, reliability, and suitability of each model for potential deployment in productive environments.

\subsubsection{LLM Runtime}
\label{sec:runtime}
A first consideration in evaluating practicality is runtime, i.e., the time it takes a model to analyse a single website.
Runtime performance is especially critical in operational settings, where large volumes of websites must be processed, and rapid decision-making is required to prevent harm caused by phishing attacks.

\begin{figure}[ht!]
    \centering
    \includegraphics[width=1\linewidth]{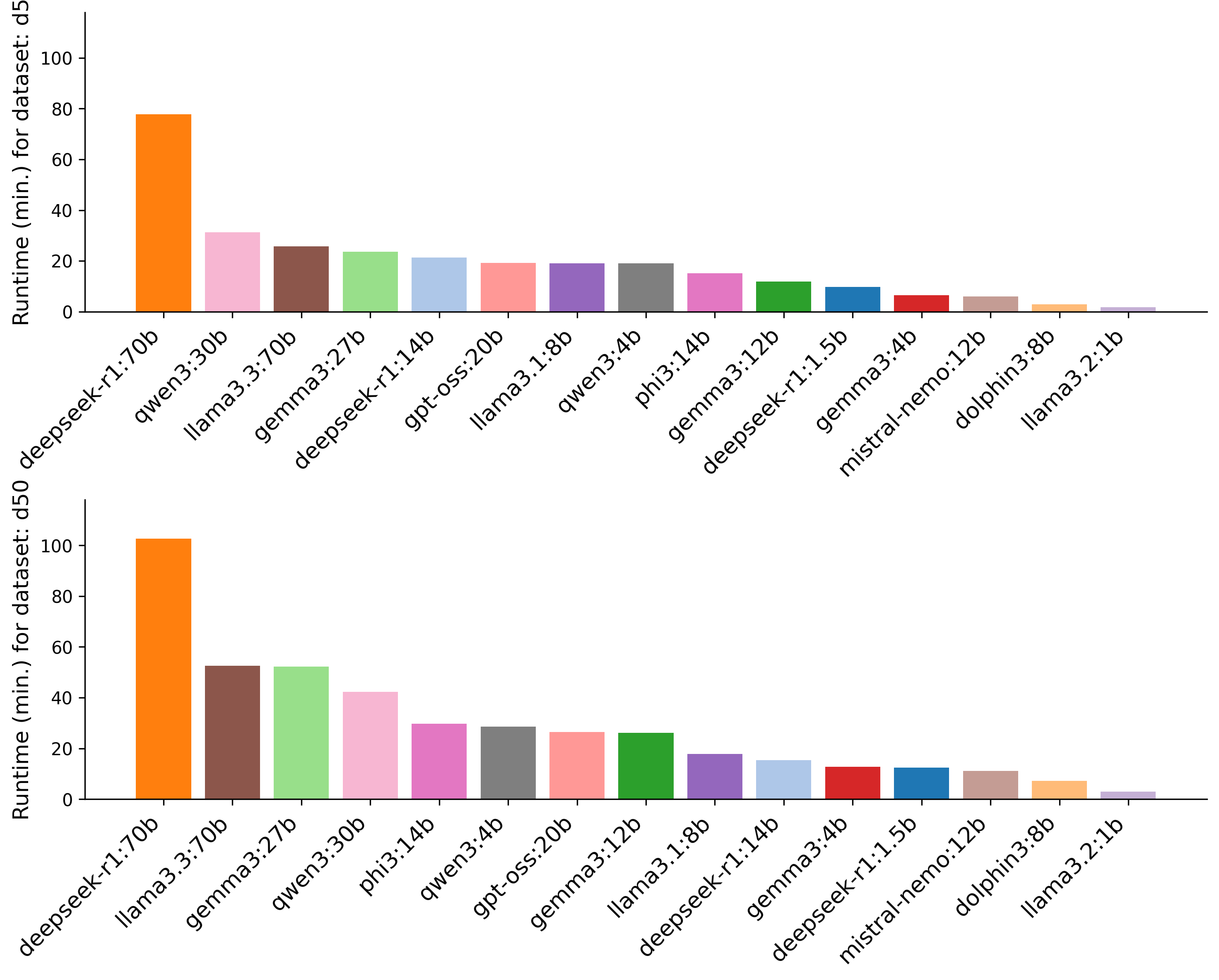}
    \caption{Runtime for each model with the D5 and D50 dataset.}
    \label{fig:runtime_hist}
\end{figure}

{\Cref{fig:runtime_hist}} shows the total runtime per model for both datasets.
While running the experiment using the A100 GPU with a 131,072k context window length, the two 70b models (llama3.3:70B and the deepseek-r1:70B) exceeded the VRAM capacity and therefore produced a runtime error.
To still be able to test these two models, a switch to the more powerful H100 GPU was made. Since the 131,072k token context window still exceeded the VRAM, we decided to reduce the window length to 65,536k tokens.

According to the Ollama model list, all models have a context window of 128k tokens. However, we set the value to 131,072k tokens, which potentially allows for longer context than the models can handle, as we found it common practice to use powers of two as a context window limit. Since no prompt used in this work exceeds 128k tokens, this has no impact on the model's performance. If a prompt were to exceed a model's context limit, Ollama automatically truncates the prompt to fit the model's abilities.

Analysing the 40 websites in the D5 dataset required a total of 291 minutes ($\sim$ 5 hours), while analysing the same websites in the D50 dataset
took 440.5 minutes ($\sim$ 7 hours and 25 minutes), which is about 51.4\% longer. 
This clearly indicates that the prompt length has a significant influence on total runtime, which is a key consideration for Experiment 2, where the dataset size increases to 1,000 websites.

As expected, larger models, such as deepseek-r1:70b and llama3.3:70b, consistently required the most time for each of the two dataset configurations, ranging from 25.8 to 102.5 minutes, whereas the smallest models, like llama3.2:1b and dolphin3:8b, completed the task within just 1.8 to 7 minutes.
In terms of runtime per analysis run, this means that deepseek-r1:70b took on average 23.3 and 30.8 seconds per analysis run for the D5 and D50 datasets, respectively. The fastest model, llama3.2:1b, managed an analysis run on average in 0.5 and 0.9 seconds for each of the two datasets. \Cref{tab:avgruntime} in the \ref{appendix} contains the mean runtime per analysis run for all models.

Interestingly, runtime was influenced not only by model size but also by model family. For instance, the Gemma and Llama models processed data significantly more efficiently compared to others, such as DeepSeek or Qwen models. 
Interestingly, these two model families (as well as the gpt-oss model) are the only models tested in this work, which are classified as \q{thinking} models on ollama. This observation suggests that reasoning models take more time to return a result, highlighting that the model architecture plays a major role in runtime performance.

\begin{figure}[ht!]
    \centering
    \includegraphics[width=1\linewidth]{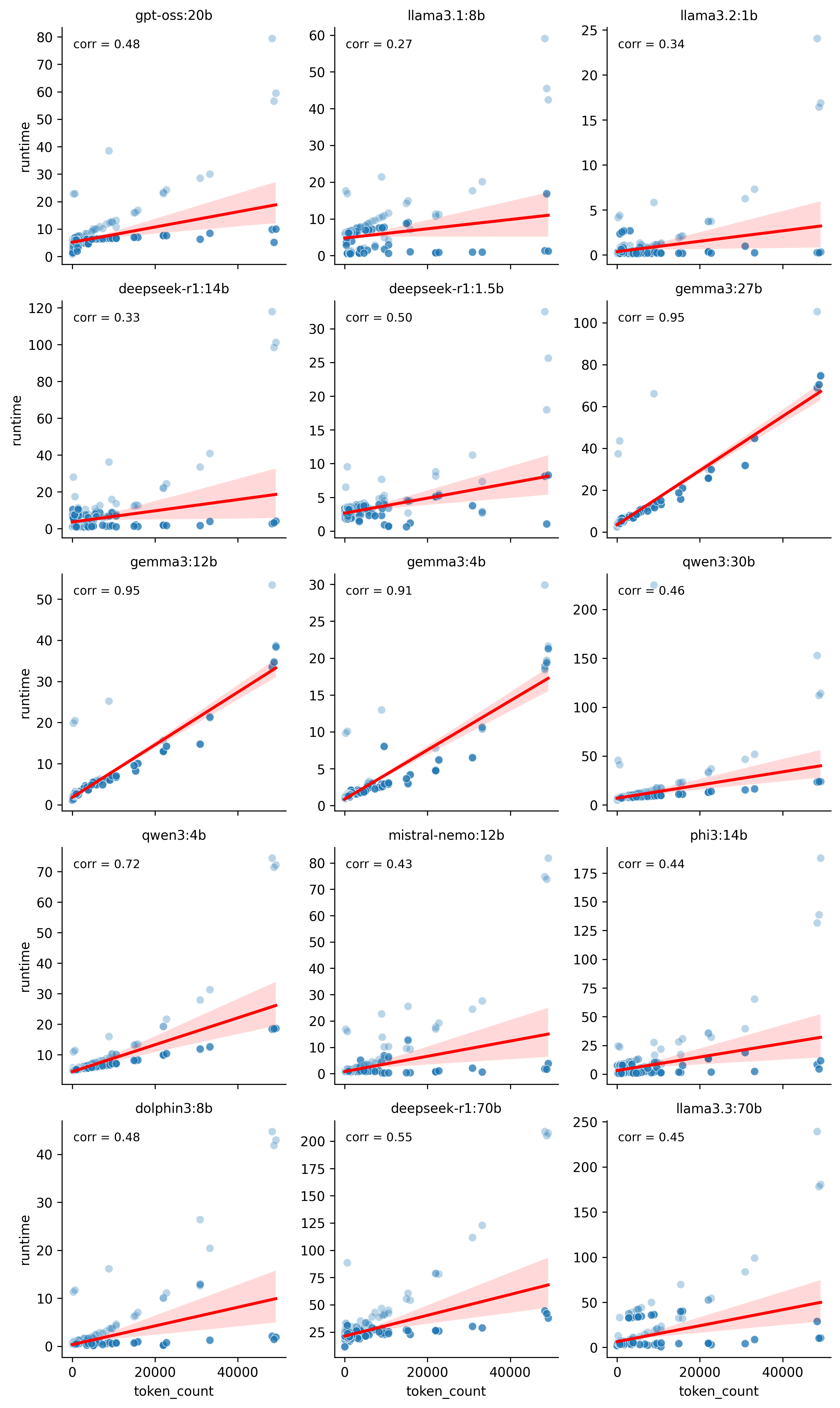}
    \caption{Correlation between model analysis runtime and prompt length. Each dot represents the token count of one website's HTML code and the corresponding analysis time. Due to websites being analysed multiple times, there is significant overlap between the 200 dots. As a consequence of the sampling scheme, there are a lot more websites towards the lower end of the token count, as they are more prevalent in the dataset.}
    \label{fig:runtime_plen_corr}
\end{figure}

An interesting observation was the significantly longer overall runtime, which indicates that the prompt length plays a crucial role in that. Potential sources of this occurrence might be the result of several factors, including the tokenisation process, as converting longer text into tokens takes more time. The self-attention mechanism scales quadratically with respect to sequence length \cite{selfatcomplex1,selfatcomplex2} and the key-value (KV) caches, which need to be built and get larger for longer prompts \cite{KVCache}.
{\Cref{fig:runtime_plen_corr}} illustrates the relationship between 
HTML token count and analysis runtime. The data includes all five analysis runs for each website across the D5 and D50 datasets for each model, totalling 400 data points. The large discrepancy of the HTML token count can be explained by the dataset sampling scheme, which exactly aimed at including websites of all lengths.
Most models show a positive correlation, with correlation coefficients ranging from 0.27 to 0.95. This suggests that, on average, all models experience an increase in runtime with longer prompts; however, the strength of this relationship varies between model families rather than being directly tied to model size. 
For example, Gemma3 or the DeepSeek models exhibit a similar correlation coefficient within their respective families, underscoring the importance of architectural design in runtime efficiency.

Another way to compare the runtime
is to compute the relative runtime per model between the two datasets. To do this, the runtime for each model and website in the D5 dataset was divided by the runtime for the same model and website in the D50 dataset. The results were then logarithmised to handle outliers and improve interpretability. 

\begin{figure}[ht!]
    \centering
    \includegraphics[width=1\linewidth]{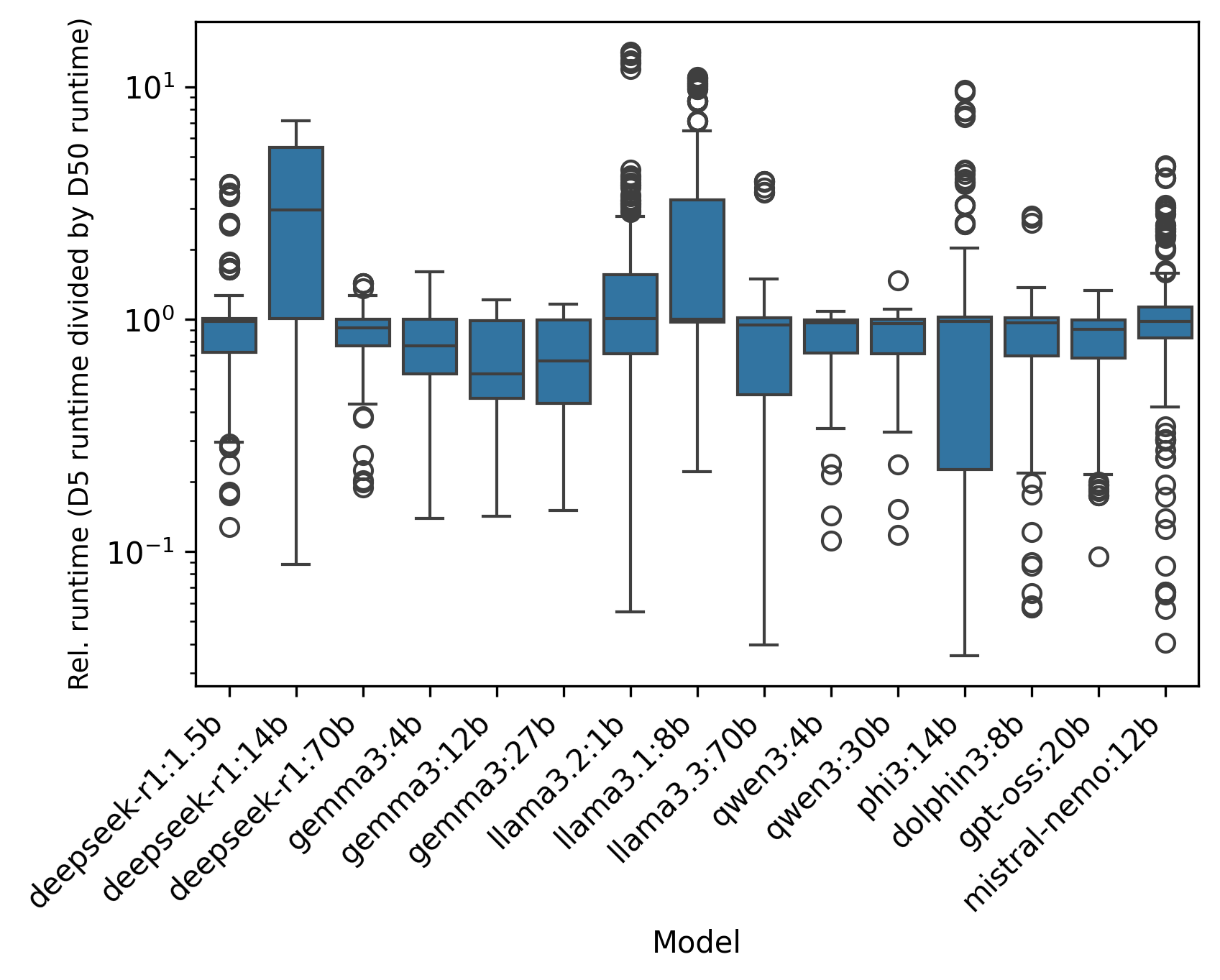}
    \caption{Distribution of the logarithm of the relative analysis time differences between the D5 and D50 datasets for each model. The distributions are based on 200 analysis runs per model, meaning that each website’s time difference is included five times (once per run). Values below 0 indicate that analysing a website with fewer HTML tokens (D5) was faster than analysing the same website with more tokens (D50).
    }
    \label{fig:runtime_bplot}
\end{figure}

{\Cref{fig:runtime_bplot}} shows the resulting boxplots.

For most models, the boxes fall below $10^0$, indicating that shorter website token counts (D5) lead to lower runtimes, as expected.
However, there are notable exceptions, including deepseek-r1:14b and llama3.1:8b models, which show reduced runtimes for longer prompts. Especially the smallest models, like the llama3.2:1b or llama3.1:8b only take a very short time (around 1 second) to analyse a website. Therefore, even small changes in analysis time can skew the results, making it take longer to process the D5 websites. The systematic behaviour of the deepseek-r1:14b and the numerous outliers of the mistral-nemo models cannot be explained as easily. Potential sources for these results are the internal model structure and optimisations, or the type of model quantisation. Especially the quantisation might be the reason why the other two deepseek models perform as expected.
The boxplots also highlight runtime variability. Some models, like the Gemma3, exhibit low runtime variability, indicated by compact boxes and a lack of outliers. In contrast, models like mistral-nemo and deepseek-r1:1.5b exhibit high variance, with outliers in both directions. 

In summary, Experiment 1 examined the runtime of the tested \Glspl{LLM}. The key findings are:
\begin{itemize}
    \item Prompt length strongly affects runtime, with longer inputs leading to higher computational costs, for almost all models.
    \item Model family, rather than model size alone, is a key determinant of runtime efficiency.
    \item Gemma and Llama models are generally faster and more consistent than DeepSeek and Qwen models.
    \item Runtime variability plays a crucial role in determining model practicality, with low-variance models being more predictable.
\end{itemize}

\subsubsection{Syntactic Performance}\label{sec:SyntPerf}
Adherence to the specified output format is another important quality of a model's reliability and usability. 
Consistent formatting allows the results to be analysed automatically and compared systematically.
In this study, models were instructed to produce their output strictly in a JSON  format (see Prompt Template ~\Cref{pr:one}), containing exactly three items: \\\texttt{is\_phishing}, \texttt{phishing\_score}, and \texttt{reasoning}. 

In practice, some models, particularly the DeepSeek family, produced the requested format but also added additional information, such as their thought process, alongside the JSON object.
An example of this behaviour is shown in Model Output 1 in ~\ref{appendix}. 
As long as the correct JSON format was present within the result, the run was counted as successful.

\begin{table}[ht]
\centering
\caption{This table shows, for each model and website, how many of the five analysis runs produced valid JSON output.
For example, the deepseek-r1:14b model generated valid JSON output in all five analysis runs for 78 websites. For one website, it never produced valid JSON output (column 0), and for another website, valid output was produced in only one run (column 1). 
In total, this covers all 80 websites (40 from the D5 dataset and 40 from the D50 dataset). The table is sorted in descending order according to column \q{5}.}
\label{tab:JSONFormatOverview}
\begin{tabular}{@{}lllllll@{}}
\toprule
\textbf{} & \multicolumn{6}{l}{\textbf{\begin{tabular}[c]{@{}l@{}}\# Runs per website with \\ correct JSON output\end{tabular}}} \\ \midrule
\multicolumn{1}{l|}{\textbf{Model Name}} & 0 & 1 & 2 & 3 & 4 & 5 \\ \midrule
\multicolumn{1}{l|}{deepseek-r1:70b} & 0 & 0 & 0 & 0 & 0 & 80 \\
\multicolumn{1}{l|}{gemma3:12b} & 0 & 0 & 0 & 0  & 0 & 80 \\
\multicolumn{1}{l|}{gemma3:27b} & 0 & 0 & 0 & 0 & 0  & 80 \\
\multicolumn{1}{l|}{deepseek-r1:14b} & 1 & 1 & 0  & 0 & 0 & 78 \\
\multicolumn{1}{l|}{gemma3:4b} & 2 & 0 & 0 & 0  & 0 & 78 \\
\multicolumn{1}{l|}{llama3.3:70b} & 2 & 1 & 0  & 0 & 1 & 76 \\
\multicolumn{1}{l|}{dolphin3:8b} & 8 & 0 & 0  & 0 & 1 & 71 \\
\multicolumn{1}{l|}{phi3:14b} & 1 & 5 & 0  & 0 & 3 & 71 \\
\multicolumn{1}{l|}{mistral-nemo:12b} & 8 & 2 & 0 & 0  & 0 & 70 \\
\multicolumn{1}{l|}{qwen3:4b} & 23 & 0 & 0 & 0  & 0 & 57 \\
\multicolumn{1}{l|}{llama3.1:8b} & 23 & 6 & 0 & 0  & 8 & 43 \\
\multicolumn{1}{l|}{deepseek-r1:1.5b} & 33 & 10 & 0 & 0  & 7 & 30 \\
\multicolumn{1}{l|}{gpt-oss:20b} & 45 & 6 & 1 & 0  & 5 & 23 \\
\multicolumn{1}{l|}{llama3.2:1b} & 77 & 0 & 0 & 0  & 1 & 2 \\
\multicolumn{1}{l|}{qwen3:30b} & 77 & 1 & 0  & 0 & 0 & 2
\end{tabular}
\end{table}


Since all model parameters were configured to produce deterministic results (see \Cref{Exp1Method}), the observed behaviour is somewhat surprising.
Under the assumption of determinism, we would expect that when analysing the same website with a given model, the output format would be consistent in all five runs---either always correct or always incorrect, which is represented by the categories 0 and 5 in {\Cref{tab:JSONFormatOverview}}.\\ 
However, some models exhibit inconsistent behaviour, producing correct JSON output only in some of the five runs (once, twice, or four times). This indicates that their output was not entirely stable.
A closer inspection of the results revealed a common pattern: in some cases, the first analysis run differed from all subsequent runs.
An example of this behaviour for the gemma3:4b model is shown in Model Output 2 in the ~\ref{appendix}.
The most plausible reason for this phenomenon is the model loading and warm-up behaviour, which includes factors such as kernel caching and background telemetry, combined with the inference framework in use, that can introduce unwanted variability \cite{ModelVar}. Since the problem cannot be easily eliminated, we decided to use a practical mitigation strategy: discard the first run and treat only the subsequent runs as true deterministic output.
An exception was observed with gpt-oss, which produced the correct JSON format in only two out of five runs for exactly one website. 
This behaviour can be attributed to the token output limit we applied in our setup.
Some models ignored the instruction and included additional explanatory text before the actual JSON output.
When the maximum token limit was reached, the output was truncated by the model, sometimes cutting off parts or all of the JSON structure, thereby rendering the result unparsable.
We decided that models that failed to produce the required output format at least four times per website across more than half of all analysed websites will be excluded from Experiment 2. 

\subsubsection{Output Stability}
\label{sec:outputstab}
In this section, we focus on output stability, specifically whether models consistently produce the same decisions across all runs. 
Each model is required to generate a JSON format output containing three items (see \Cref{sec:prompt}): \texttt{is\_phishing}, \texttt{phishing\_score}, and \texttt{reasoning}.
Therefore, in cases where the model delivered the required format, the variables were extracted and compared to determine if the information was consistent.
Ensuring coherent output is a key criterion for evaluating a model's reliability and is essential for building trust in model decisions. There are two desirable scenarios:

\begin{itemize}
    \item \texttt{is\_phishing = True}, \texttt{phishing\_score $\geq$ 5}, reasoning says it is phishing,
    \item \texttt{is\_phishing = False}, \texttt{phishing\_score $<$ 5}, reasoning says it is not phishing.
\end{itemize}

Since the phishing score can take on eleven possible values (0-10), it is unclear whether a score of five should be considered phishing detection or not. 
In cybersecurity-related tasks, it is often unclear which type of error (missing a phishing attempt or falsely flagging a benign website) is worse, as the former can potentially be costly when an intrusion occurs. The latter disrupts the user experience \cite{tnfntradeoff, intro7, type2error}.

We prioritise detecting phishing websites, as we deem the cost of missing an attack, especially in critical infrastructure, to be higher than incorrectly blocking a benign site. Therefore, a \Gls{LLM} score of five is considered as \texttt{True} when translating it into a binary classification. 

In our analysis, we analyse coherence from two complementary perspectives: i) a quantitative consistency analysis, examining the alignment between the binary decision (\texttt{is\_phishing}) and the certainty score (\texttt{phishing\_score}), and ii) qualitative coherence, examining the reasoning texts, with respect to the occurrence of certain key words that are often connected to the classification into phishing and benign websites. 
The quantitative part provides a direct measure of consistency within the result of one website analysis, ensuring alignment between the numeric outputs. 
The qualitative part focuses on the frequency of phishing-related and legitimacy-related words and phrases, as well as their negations, which captures how models articulate their decision-making and provides insight into explanation strategies and their plausibility.\\

\noindent
\underline{Quantitative Analysis}\\
The results of the quantitative consistency analysis can be found in {\Cref{tab:quanti_coher}}, which aims to highlight the relation between the two quantitative phishing measures - the boolean decision and the phishing score.

\begin{table}
\caption{This table shows the phishing scores and the classification decision (Phishing Cat) assigned by the models across the five analysis runs per website (200 runs per dataset).
Each model and dataset combination has three possible classification decisions (\texttt{True}, \texttt{False}, \texttt{NaN}). Each decision is displayed as a separate row. In cases where models did not include any of the decisions, the corresponding row is omitted. 
The three rightmost columns display the number of times no JSON output was returned (NaN) and summarise the assigned scores in two groups: $<$5 and $\geq 5$.
In \textcolor{red}{red} we highlight where the score does not match the decision.
}
\label{tab:quanti_coher}
\resizebox{\columnwidth}{!}{%
\begin{tabular}{lll|rrrrrrrrrrr|rrr}
\toprule
 & & & \multicolumn{11}{c|}{\textbf{Phishing Score}} & & & \\
\cmidrule(lr){4-14}
\textbf{Dataset} & \textbf{Model} & \textbf{Phishing Cat} & 0 & 1 & 2 & 3 & 4 & 5 & 6 & 7 & 8 & 9 & 10 & \textbf{NaN} & \textbf{$<$5} & \textbf{$\geq$5} \\
\midrule
\rowcolor{gray!10}
d5 & deepseek-r1:1.5b & False & 31 & 0 & 0 & 45 & 0 & \textcolor{red}{8} & 0 & 0 & 0 & 0 & 0 & 0 & 76 & 8 \\
\rowcolor{gray!10}
d5 & deepseek-r1:1.5b & True  & 0 & 0 & 0 & 0 & 0 & 0 & 0 & 5 & 0 & 30 & 0 & 0 & 0 & 35 \\
\rowcolor{gray!10}
d5 & deepseek-r1:1.5b & NaN   & 0 & 0 & 0 & 0 & 0 & 0 & 0 & 0 & 0 & 0 & 0 & 81 & 0 & 0 \\
d5 & deepseek-r1:14b & False & 56 & 9 & 10 & 73 & 1 & 0 & 0 & 0 & 0 & 0 & 0 & 0 & 149 & 0 \\
d5 & deepseek-r1:14b & True  & 0 & 0 & 0 & 0 & 0 & 23 & 0 & 19 & 5 & 0 & 0 & 0 & 0 & 47 \\
d5 & deepseek-r1:14b & NaN   & 0 & 0 & 0 & 0 & 0 & 0 & 0 & 0 & 0 & 0 & 0 & 4 & 0 & 0 \\
\rowcolor{gray!10}
d5 & deepseek-r1:70b & False & 4 & 64 & 43 & 28 & 6 & 0 & 0 & 0 & 0 & 0 & 0 & 0 & 145 & 0 \\
\rowcolor{gray!10}
d5 & deepseek-r1:70b & True  & 0 & 0 & 0 & 0 & 0 & 14 & 5 & 17 & 17 & 2 & 0 & 0 & 0 & 55 \\
d5 & dolphin3:8b     & False & 10 & 5 & 76 & 34 & 0 & 0 & 0 & 0 & 0 & 0 & 0 & 0 & 125 & 0 \\
d5 & dolphin3:8b & True & 0 & 0 & 0 & 0 & 0 & 10 & 25 & 30 & 0 & 0 & 0 & 0 & 0 & 65 \\
d5 & dolphin3:8b & NaN & 0 & 0 & 0 & 0 & 0 & 0 & 0 & 0 & 0 & 0 & 0 & 10 & 0 & 0 \\
\rowcolor{gray!10}
d5 & gemma3:12b & False & 0 & 20 & 20 & 60 & 5 & 0 & 0 & 0 & 0 & 0 & 0 & 0 & 105 & 0 \\
\rowcolor{gray!10}
d5 & gemma3:12b & True & 0 & 0 & 0 & 0 & 0 & 0 & 65 & 25 & 5 & 0 & 0 & 0 & 0 & 95 \\
d5 & gemma3:27b & False & 0 & 0 & 0 & 35 & 35 & \textcolor{red}{30} & 0 & 0 & 0 & 0 & 0 & 0 & 70 & 30 \\
d5 & gemma3:27b & True & 0 & 0 & 0 & 0 & 0 & 5 & 40 & 45 & 10 & 0 & 0 & 0 & 0 & 100 \\
\rowcolor{gray!10}
d5 & gemma3:4b & False & 0 & 0 & 0 & 25 & 0 & 0 & 0 & 0 & 0 & 0 & 0 & 0 & 25 & 0 \\
\rowcolor{gray!10}
d5 & gemma3:4b & True & 0 & 0 & 0 & 0 & 0 & 5 & 40 & 90 & 35 & 0 & 0 & 0 & 0 & 170 \\
\rowcolor{gray!10}
d5 & gemma3:4b & NaN & 0 & 0 & 0 & 0 & 0 & 0 & 0 & 0 & 0 & 0 & 0 & 5 & 0 & 0 \\
d5 & gpt-oss:20b & False & 15 & 48 & 8 & 8 & 0 & 0 & 0 & 0 & 0 & 0 & 0 & 0 & 79 & 0 \\
d5 & gpt-oss:20b & True & 0 & 0 & 0 & 0 & 0 & 4 & 0 & 0 & 7 & 2 & 0 & 0 & 0 & 13 \\
d5 & gpt-oss:20b & NaN & 0 & 0 & 0 & 0 & 0 & 0 & 0 & 0 & 0 & 0 & 0 & 108 & 0 & 0 \\
\rowcolor{gray!10}
d5 & llama3.1:8b & False & 6 & 0 & 14 & 1 & 0 & 0 & 0 & 0 & 0 & 0 & 0 & 0 & 21 & 0 \\
\rowcolor{gray!10}
d5 & llama3.1:8b & True & 0 & 0 & 0 & 0 & 0 & 9 & 0 & 0 & 34 & 44 & 0 & 0 & 0 & 87 \\
\rowcolor{gray!10}
d5 & llama3.1:8b & NaN & 0 & 0 & 0 & 0 & 0 & 0 & 0 & 0 & 0 & 0 & 0 & 92 & 0 & 0 \\
d5 & llama3.2:1b & False & 0 & 0 & 0 & 0 & 0 & 0 & 5 & 0 & 0 & 0 & 0 & 0 & 0 & 5 \\
d5 & llama3.2:1b & NaN & 0 & 0 & 0 & 0 & 0 & 0 & 0 & 0 & 0 & 0 & 0 & 195 & 0 & 0 \\
\rowcolor{gray!10}
d5 & llama3.3:70b & False & 5 & 14 & 101 & 0 & 0 & 0 & 0 & 0 & 0 & 0 & 0 & 0 & 120 & 0 \\
\rowcolor{gray!10}
d5 & llama3.3:70b & True & 0 & 0 & 0 & 0 & 0 & 0 & 5 & 0 & 66 & 9 & 0 & 0 & 0 & 80 \\
d5 & mistral-nemo:12b & False & 0 & 0 & 33 & 51 & 1 & \textcolor{red}{4} & \textcolor{red}{6} & 0 & 0 & 0 & 0 & 0 & 85 & 10 \\
d5 & mistral-nemo:12b & True & 0 & 0 & 0 & 0 & 0 & 5 & 8 & 32 & 22 & 33 & 0 & 0 & 0 & 100 \\
d5 & mistral-nemo:12b & NaN & 0 & 0 & 0 & 0 & 0 & 0 & 0 & 0 & 0 & 0 & 0 & 5 & 0 & 0 \\
\rowcolor{gray!10}
d5 & phi3:14b & False & 84 & 6 & 21 & 5 & 0 & 0 & 0 & 0 & 0 & 0 & 0 & 0 & 116 & 0 \\
\rowcolor{gray!10}
d5 & phi3:14b & True & 0 & 0 & 0 & 0 & 0 & 5 & 4 & 37 & 15 & 10 & 0 & 0 & 0 & 71 \\
\rowcolor{gray!10}
d5 & phi3:14b & NaN & 0 & 0 & 0 & 0 & 0 & 0 & 0 & 0 & 0 & 0 & 0 & 13 & 0 & 0 \\
d5 & qwen3:30b & False & 5 & 4 & 0 & 0 & 0 & 0 & 0 & 0 & 0 & 0 & 0 & 0 & 9 & 0 \\
d5 & qwen3:30b & True & 0 & 0 & 0 & 0 & 0 & 1 & 0 & 0 & 1 & 0 & 0 & 0 & 0 & 2 \\
d5 & qwen3:30b & NaN & 0 & 0 & 0 & 0 & 0 & 0 & 0 & 0 & 0 & 0 & 0 & 189 & 0 & 0 \\
\rowcolor{gray!10}
d5 & qwen3:4b & False & 95 & 0 & 0 & 0 & 0 & 0 & 0 & 0 & 0 & 0 & 0 & 0 & 95 & 0 \\
\rowcolor{gray!10}
d5 & qwen3:4b & NaN & 0 & 0 & 0 & 0 & 0 & 0 & 0 & 0 & 0 & 0 & 0 & 105 & 0 & 0 \\
\midrule
d50 & deepseek-r1:1.5b & False & 1 & 0 & 0 & 4 & 0 & \textcolor{red}{12} & 0 & 0 & 0 & 0 & 0 & 0 & 5 & 12 \\
d50 & deepseek-r1:1.5b & True & \textcolor{red}{4} & 0 & 0 & 0 & 0 & 1 & 0 & 11 & 0 & 36 & 0 & 0 & 4 & 48 \\
d50 & deepseek-r1:1.5b & NaN & 0 & 0 & 0 & 0 & 0 & 0 & 0 & 0 & 0 & 0 & 0 & 131 & 0 & 0 \\
\rowcolor{gray!10}
d50 & deepseek-r1:14b & False & 50 & 0 & 0 & 79 & 0 & 0 & 0 & 0 & 0 & 0 & 0 & 0 & 129 & 0 \\
\rowcolor{gray!10}
d50 & deepseek-r1:14b & True & 0 & 0 & 0 & 0 & 0 & 31 & 0 & 21 & 14 & 0 & 0 & 0 & 0 & 66 \\
\rowcolor{gray!10}
d50 & deepseek-r1:14b & NaN & 0 & 0 & 0 & 0 & 0 & 0 & 0 & 0 & 0 & 0 & 0 & 5 & 0 & 0 \\
d50 & deepseek-r1:70b & False & 11 & 78 & 31 & 0 & 14 & 0 & 0 & 0 & 0 & 0 & 0 & 0 & 134 & 0 \\
d50 & deepseek-r1:70b & True & 0 & 0 & 0 & 0 & 0 & 12 & 4 & 30 & 19 & 1 & 0 & 0 & 0 & 66 \\
\rowcolor{gray!10}
d50 & dolphin3:8b & False & 0 & 5 & 29 & 20 & 0 & 0 & 0 & 0 & 0 & 0 & 0 & 0 & 54 & 0 \\
\rowcolor{gray!10}
d50 & dolphin3:8b & True & 0 & 0 & 0 & 0 & 0 & 26 & 48 & 41 & 0 & 0 & 0 & 0 & 0 & 115 \\
\rowcolor{gray!10}
d50 & dolphin3:8b & NaN & 0 & 0 & 0 & 0 & 0 & 0 & 0 & 0 & 0 & 0 & 0 & 31 & 0 & 0 \\
d50 & gemma3:12b & False & 0 & 10 & 5 & 40 & 15 & 0 & 0 & 0 & 0 & 0 & 0 & 0 & 70 & 0 \\
d50 & gemma3:12b & True & 0 & 0 & 0 & 0 & 0 & 0 & 65 & 45 & 20 & 0 & 0 & 0 & 0 & 130 \\
\rowcolor{gray!10}
d50 & gemma3:27b & False & 0 & 0 & 0 & 15 & 45 & \textcolor{red}{10} & \textcolor{red}{15} & 0 & 0 & 0 & 0 & 0 & 60 & 25 \\
\rowcolor{gray!10}
d50 & gemma3:27b & True & 0 & 0 & 0 & 0 & 0 & 0 & 35 & 50 & 30 & 0 & 0 & 0 & 0 & 115 \\
d50 & gemma3:4b & True & 0 & 0 & 0 & 0 & 0 & 0 & 40 & 70 & 85 & 0 & 0 & 0 & 0 & 195 \\
d50 & gemma3:4b & NaN & 0 & 0 & 0 & 0 & 0 & 0 & 0 & 0 & 0 & 0 & 0 & 5 & 0 & 0 \\
\rowcolor{gray!10}
d50 & gpt-oss:20b & False & 5 & 27 & 5 & 4 & 0 & 0 & 0 & 0 & 0 & 0 & 0 & 0 & 41 & 0 \\
\rowcolor{gray!10}
d50 & gpt-oss:20b & True & 0 & 0 & 0 & 0 & 0 & 0 & 0 & 0 & 4 & 6 & 0 & 0 & 0 & 10 \\
\rowcolor{gray!10}
d50 & gpt-oss:20b & NaN & 0 & 0 & 0 & 0 & 0 & 0 & 0 & 0 & 0 & 0 & 0 & 149 & 0 & 0 \\
d50 & llama3.1:8b & False & 10 & 0 & 4 & 0 & 1 & 0 & 0 & 0 & 0 & 0 & 0 & 0 & 15 & 0 \\
d50 & llama3.1:8b & True & 0 & 0 & 0 & 0 & 0 & 5 & 0 & 1 & 81 & 39 & \textcolor{red}{4\footnotemark} & 0 & 0 & 130 \\
d50 & llama3.1:8b & NaN & 0 & 0 & 0 & 0 & 0 & 0 & 0 & 0 & 0 & 0 & 0 & 55 & 0 & 0 \\
\rowcolor{gray!10}
d50 & llama3.2:1b & False & 0 & 0 & 0 & 0 & 0 & 0 & \textcolor{red}{5} & 0 & 0 & 0 & 0 & 0 & 0 & 5 \\
\rowcolor{gray!10}
d50 & llama3.2:1b & True & 0 & 0 & 0 & 0 & 0 & 0 & 0 & 0 & 4 & 0 & 0 & 0 & 0 & 4 \\
\rowcolor{gray!10}
d50 & llama3.2:1b & NaN & 0 & 0 & 0 & 0 & 0 & 0 & 0 & 0 & 0 & 0 & 0 & 191 & 0 & 0 \\
d50 & llama3.3:70b & False & 5 & 5 & 59 & 0 & 0 & \textcolor{red}{1} & \textcolor{red}{5} & 0 & 0 & 0 & 0 & 0 & 69 & 6 \\
d50 & llama3.3:70b & True & 0 & 0 & 0 & 0 & 0 & 0 & 0 & 0 & 101 & 9 & 0 & 0 & 0 & 110 \\
d50 & llama3.3:70b & NaN & 0 & 0 & 0 & 0 & 0 & 0 & 0 & 0 & 0 & 0 & 0 & 15 & 0 & 0 \\
\rowcolor{gray!10}
d50 & mistral-nemo:12b & False & 5 & 0 & 15 & 16 & 0 & \textcolor{red}{4} & 0 & 0 & 0 & 0 & 0 & 0 & 36 & 4 \\
\rowcolor{gray!10}
d50 & mistral-nemo:12b & True & 0 & 0 & 0 & 0 & 0 & 0 & 1 & 42 & 39 & 35 & 0 & 0 & 0 & 117 \\
\rowcolor{gray!10}
d50 & mistral-nemo:12b & NaN & 0 & 0 & 0 & 0 & 0 & 0 & 0 & 0 & 0 & 0 & 0 & 43 & 0 & 0 \\
d50 & phi3:14b & False & 19 & 5 & 5 & 14 & 0 & 0 & 0 & 0 & 0 & 0 & 0 & 0 & 43 & 0 \\
d50 & phi3:14b & True & 0 & 0 & 0 & 0 & 0 & 4 & 0 & 116 & 12 & 10 & 0 & 0 & 0 & 142 \\
d50 & phi3:14b & NaN & 0 & 0 & 0 & 0 & 0 & 0 & 0 & 0 & 0 & 0 & 0 & 15 & 0 & 0 \\
\rowcolor{gray!10}
d50 & qwen3:30b & NaN & 0 & 0 & 0 & 0 & 0 & 0 & 0 & 0 & 0 & 0 & 0 & 200 & 0 & 0 \\
d50 & qwen3:4b & False & 190 & 0 & 0 & 0 & 0 & 0 & 0 & 0 & 0 & 0 & 0 & 0 & 190 & 0 \\
d50 & qwen3:4b & NaN & 0 & 0 & 0 & 0 & 0 & 0 & 0 & 0 & 0 & 0 & 0 & 10 & 0 & 0 \\
\bottomrule
\end{tabular}

}

\end{table}
\footnotetext[13]{The model assigned a score of 12.}

The model output shows coherence in almost all cases between the boolean decision and the assigned score. 
Almost all inconsistencies can be attributed to cases where the boolean decision was \texttt{False} while the corresponding score was 5. 
An exception was observed with the Mistral model, which produced a different type of inconsistency by assigning a score of 6 to the decision \texttt{False}.
The \texttt{NaN} rows and columns indicate instances where the model failed to produce a valid JSON output.

Notably, no model ever assigned the highest phishing score of 10, while the lowest score of 0 occurred frequently across models. In four cases, the llama3.1:8b model determined that twelve is an appropriate score, which, of course, is considered an error, as the score should be ten or lower.
The tendency of the models to assign the lowest available scores and never the highest possible score indicates that models might find it easier to conclude when a website is benign based on the absence of phishing indicators. This behaviour could also be influenced by the prompt design chosen, where models check all the criteria of the provided examples, and if none are met, assign a low score. Another reason this result might appear is that phishing websites often contain elements that can also be found in benign websites. This can lead to the model being less certain when deciding on phishing, as indicators from both benign and malicious websites are present simultaneously.

Overall, the models showed a very similar behaviour when giving scores for the respective boolean decision, as well as the amount of missing JSON outputs (NaNs). This indicates that using longer HTML code does not guarantee a different (or better) classification result. 

The few exceptions are the gemma3:4b model, which assigned significantly more scores $<$5 for the D5 dataset compared to the D50 dataset, and the gpt-oss:20b, llama3.2:1b and the qwen3:4b models, where the number on NaN outputs was noticeably smaller for the D50 dataset. Overall, the results suggest that the D5 dataset provides sufficient data for a reasonable analysis.

Diving deeper into the dataset differences reveals that most models retain their assessment for the same website, as shown in {\Cref{tab:ds_diff}}. Eight models score 30 or above (out of 40 possible websites) in column \q{0}, which means that for 75\% or more of the websites, the models produced the same boolean result in all 5 analysis runs, regardless of the dataset. Mainly, the larger tested models stay consistent between the two datasets, with the notable exception being the llama3.2:1b model. However, this model produced almost exclusively NaN values in both cases.

The second most common case is that models changed all five of their assessments, shown in column \q{5}. According to {\Cref{tab:JSONFormatOverview}}, the number of NaN classifications for the concerning models changes notably between the two datasets. This indicates that models did not change their opinion but rather that the changes are mostly related to the output being JSON or not.

The remaining observed differences, denoted in columns \q{1} to \q{4}, are related to the fact that the output of some models only becomes consistent after the first analysis run.\\

\begin{table}
\centering
\caption{Overview of the classification consistency across models for the D5 and D50 analyses. For each website, the number of differing classifications across five runs is counted. A value of 0 indicates identical results in all runs, while a value of 5 means that each run produced different results. The counts are then aggregated over all 40 websites for each model. The models are listed in descending order according to column \q{0}.}

\label{tab:ds_diff}
\begin{tabular}{@{}lllllll@{}}
\toprule
\textbf{} & \multicolumn{6}{l}{\textbf{\begin{tabular}[c]{@{}l@{}}\# Classification result differences \\ between the D5 and D50 dataset \\ per website runs\end{tabular}}} \\ \midrule
\multicolumn{1}{l|}{\textbf{Model Name}} & 0 & 1 & 2 & 3 & 4 & 5 \\ \midrule
\multicolumn{1}{l|}{llama3.2:1b} & 39 & 0 & 0 & 0 & 1 & 0 \\
\multicolumn{1}{l|}{qwen3:30b} & 37 & 1 & 0 & 0 & 0 & 2 \\
\multicolumn{1}{l|}{gemma3:27b} & 35 & 0 & 0 & 0 & 0 & 5 \\
\multicolumn{1}{l|}{gemma3:4b} & 33 & 0 & 0 & 0 & 0 & 7 \\
\multicolumn{1}{l|}{deepseek-r1:14b} & 30 & 1 & 0 & 0 & 2 & 7 \\
\multicolumn{1}{l|}{deepseek-r1:70b} & 30 & 1 & 0 & 0 & 3 & 6 \\
\multicolumn{1}{l|}{gemma3:12b} & 29 & 0 & 0 & 0 & 0 & 11 \\
\multicolumn{1}{l|}{llama3.3:70b} & 29 & 1 & 0 & 0 & 2 & 8 \\
\multicolumn{1}{l|}{dolphin3:8b} & 24 & 1 & 0 & 0 & 0 & 15 \\
\multicolumn{1}{l|}{mistral-nemo:12b} & 24 & 0 & 0 & 0 & 2 & 14 \\
\multicolumn{1}{l|}{qwen3:4b} & 21 & 0 & 0 & 0 & 0 & 19 \\
\multicolumn{1}{l|}{gpt-oss:20b} & 20 & 4 & 0 & 1 & 4 & 11 \\
\multicolumn{1}{l|}{phi3:14b} & 19 & 3 & 0 & 0 & 3 & 15 \\
\multicolumn{1}{l|}{llama3.1:8b} & 16 & 7 & 0 & 0 & 4 & 13 \\
\multicolumn{1}{l|}{deepseek-r1:1.5b} & 13 & 5 & 0 & 0 & 4 & 18
\end{tabular}
\end{table}

\noindent
\underline{Qualitative Analysis of Reasoning Outputs}\\
To complement the coherence checks, we conducted a qualitative analysis of the models’ reasoning outputs, i.e., the explanatory texts contained in the \texttt{reasoning} key of the JSON output. 
The analysis focuses on the language used by the models to justify their decisions. Specifically, we scanned the reasoning texts for the occurrence of predefined keywords: (i) phishing indicators, terms associated with malicious or suspicious behaviour (e.g. \q{phishing}, \q{suspicious}, \q{fraudulent}, \q{deceptive}), and (ii) legitimacy indicators, terms associated with safe or trusted websites (e.g. \q{legitimate}, \q{safe}, \q{no phishing indicators}, \q{trusted}). These keywords were derived by manually analysing model outputs during the experimentation phase.
For each model–dataset pair, we counted the total mentions of each category and reported them as \texttt{phishing\_score\_tendency} and \texttt{legit\_score\_tendency}, respectively. 
While this keyword-based approach is inherently high-level and depends on the defined keyword set, it provides an indication of the style and focus of the models' explanation.

The results, summarised in {\Cref{tab:quali_ana}}, which can be found in ~\ref{appendix}, reveal differences in how models frame their reasoning. Some models, such as gemma3:4b, use many terms that we, based on our list, would associate with a phishing website. This result is also in line with the observation that the gemma3:4b model assigned almost all websites to be phishing. 
Others, such as mistral-nemo, return a more balanced amount of phrases and keywords associated with phishing or benign websites, which is again aligned with a more balanced boolean decision profile. We also observe variability across datasets: for instance, dolphin3:8b uses more language associated with phishing websites on the d5 dataset compared to the d50 dataset. 

\subsubsection{Consequences for the setup of Experiment 2}
The results from Experiment 1 were instrumental in determining an efficient and reliable setup for Experiment 2. 
We excluded the following models, due to their failures to generate the required JSON result format: llama3.1:8b, llama3.2:1b, gpt-oss:20b, qwen3:4b, qwen3:30b and deepseek.r1:1.5b.

Furthermore, we observed that models are consistent after the first analysis run~\Cref{sec:SyntPerf}. To account for this, only two runs per website will be conducted. The first run serves as initiation and will be discarded, while the second run will be used as final decision output.

To maintain reasonable runtimes analysing the larger dataset, all websites will be truncated using the D5 truncation scheme, retaining at most 5\% of the original HTML code for analysis. 
This choice is supported by the quantitative coherence check, which showed minimal performance differences between the D5 and the D50 datasets.

Finally, while the context window size could be reduced due to the smaller token number in the D5 scheme, 
we decided against modifying model parameters, as doing so could negatively impact comparability between the experiments.

\subsection{Experiment 2}
\label{sec:ex2res}
In this experiment, the objective is to evaluate the classification performance of the models on a larger dataset comprising 1,000 websites. 
Performance is measured using well-established evaluation metrics, namely accuracy, precision, recall, and F1-score.

\softbox{
  \textbf{Accuracy = } $\frac{TP+TN}{P+N}$  

  \textbf{Precision = }$\frac{TP}{TP + FP}$

  \textbf{Recall = } $\frac{TP}{TP + FN}$

  \textbf{F1-score = } $2* \frac{Precision * Recall}{Precision + Recall}$

  \textbf{where:} P = Positive, N = Negative, TP = True Positive, TN = True Negative,\\
  FP = False Positive, FN = False Negatives
}

{\Cref{tab:perf_metrics}} and {\Cref{tab:perf_metrics_num}} show the performance results.

For {\Cref{tab:perf_metrics}}, the boolean decision was taken as the predicted label, whereas {\Cref{tab:perf_metrics_num}} shows the metrics based on the numeric decision of the models. Numeric values smaller than five correspond to a decision of \q{not phishing} and all values of five or larger correspond to the decision \q{phishing}.

\begin{table}
    \caption{Summary of performance metrics per model based on the boolean classification decision, sorted in descending order by F1 score.}
    \label{tab:perf_metrics}
    \resizebox{\columnwidth}{!}{%
    \begin{tabular}{lrrrrr}
    \toprule
    \textbf{Model} & \textbf{F1\_score} & \textbf{Accuracy} & \textbf{Precision} & \textbf{Recall} & \textbf{NaN Proportion} \\
    \midrule
    llama3.3:70b & \textbf{0.893} & \textbf{0.887} & 0.845 & 0.948 & 0.001 \\
    deepseek-r1:70b & 0.873 & 0.865 & 0.824 & 0.929 & 0.073 \\
    mistral-nemo:12b & 0.858 & 0.849 & 0.811 & 0.909 & 0.009 \\
    deepseek-r1:14b & 0.842 & 0.842 & 0.810 & 0.877 & 0.086 \\
    gemma3:27b & 0.835 & 0.809 & 0.737 & 0.964 & 0.004 \\
    gemma3:12b & 0.828 & 0.804 & 0.734 & 0.951 & 0.008 \\
    gemma3:4b & 0.690 & 0.562 & 0.533 & \textbf{0.980} & 0.007 \\
    dolphin3:8b & 0.651 & 0.729 & \textbf{0.913} & 0.506 & 0.000 \\
    phi3:14b & 0.462 & 0.601 & 0.701 & 0.345 & 0.014 \\
    \bottomrule
    \end{tabular}
    }
\end{table}

\begin{table}
    \caption{Summary of performance metrics per model based on the numeric classification decision, sorted in descending order by F1 score.}
    \label{tab:perf_metrics_num}
    \resizebox{\columnwidth}{!}{%
    \begin{tabular}{lrrrrr}
    \toprule
     \textbf{Model} & \textbf{F1\_score} & \textbf{Accuracy} & \textbf{Precision} & \textbf{Recall} & \textbf{NaN Proportion} \\
    \midrule
    llama3.3:70b & \textbf{0.892} & \textbf{0.885} & 0.842 & 0.948 & 0.001 \\
    deepseek-r1:70b & 0.870 & 0.862 & 0.821 & 0.927 & 0.073 \\
    mistral-nemo:12b & 0.843 & 0.823 & 0.761 & 0.946 & 0.009 \\
    deepseek-r1:14b & 0.842 & 0.839 & 0.795 & 0.895 & 0.086 \\
    gemma3:12b & 0.828 & 0.803 & 0.732 & 0.951 & 0.008 \\
    gemma3:27b & 0.764 & 0.690 & 0.618 & \textbf{1.000} & 0.004 \\
    dolphin3:8b & 0.651 & 0.729 & \textbf{0.913} & 0.506 & 0.000 \\
    phi3:14b & 0.462 & 0.601 & 0.701 & 0.345 & 0.014 \\
    \bottomrule
    \end{tabular}
    }
\end{table}

The evaluation in {\Cref{tab:perf_metrics}}, based on the boolean decision output, highlights the relative strengths of the different models in phishing website classification. When prioritising overall effectiveness, the F1-score and accuracy are particularly informative, as both false positives and false negatives can have serious consequences in this domain.

Llama3.3:70b emerges as the strongest overall model, achieving the highest F1-score (0.893) and accuracy (0.887). It maintains a good balance between precision (0.845) and recall
(0.948), with an almost negligible proportion of NaNs (0.001), making it highly reliable for deployment. Similarly, deepseek-r1:70b achieves a strong F1-score (0.873) and accuracy (0.865), balancing precision (0.824) and recall (0.929), while maintaining consistent output formatting (NaN proportion: 0.073).

Mistral-nemo:12b also performs strongly, with an F1-score of 0.858 and accuracy of 0.849. It emphasises a balanced trade-off between precision (0.811) and recall (0.909), and a low rate of
missing predictions (0.009). The best-performing models all exhibit higher recall than precision, indicating greater success in identifying actual phishing sites than in correctly identifying benign websites.

Mistral-nemo:12b, deepseek-r1:14b, gemma3:27b and gemma3:12b show solid all-round performance (F1-score 0.828-0.858, accuracy 0.804-0.865), and similar error patterns compared to the top models. While the overall performance is slightly worse than for the 70b models, this tier of models is considerably smaller, which affects deployability and analysis runtime.

At the lower end, gemma3:4b, dolphin3:8b, and phi3:14b show the weakest results, with F1-Scores (0.462-0.690) and accuracies (0.562-0.729).

Overall, models such as llama3.3:70b and deepseek-r1:70b provide the most balanced and reliable performance, excelling in both F1-score and accuracy. Mistral-nemo:12b and gemma3:27b also perform strongly when maximising recall is essential, while deepseek-r1:14b offers dependable balanced detection. Overall, model size impacts the result performance, with the largest models performing the best. However, considerably smaller models only perform marginally worse. 

The results in {\Cref{tab:perf_metrics_num}} present a similar picture, where instead of the boolean decision, the numeric categorisation was taken as the decision criterion. 
Consistent with Experiment 1, we observe that models only rarely assign conflicting Boolean decisions and numeric assessments. Overall, however, measuring the model's performance based on the phishing score results in a slightly worse performance across most performance metrics in almost all cases.
A notable outlier is gemma3:27b, which now achieves only a 0.764 F1-score, down from 0.835 for the Boolean decision.
In conclusion, the models' binary decisions yield better results across almost all tested models, indicating that relying solely on a binary decision can be sufficient.

In Section \ref{sec:relW}, the results of comparable approaches are summarised. Most prior studies report results using large, proprietary models, such as Gemini Pro 1.0 or GPT-4, with F1-scores reaching up to 0.99. However, due to differences in the analysis architecture and datasets, these results should be interpreted with caution, as direct comparisons are not always meaningful. Overall, the evidence suggests that large models tend to outperform smaller ones, which is consistent with expectations. Nevertheless, it is encouraging to see that the tested 70b models come close to 90\% F1-score and accuracy, while more compact models sit around 85\% for both metrics. This shows the improvement in small open-source models, which previously achieved only around 74\% accuracy or were worse, for state-of-the-art models one or two years ago~\cite{propBM1}. This suggests that smaller, more efficient models are closing the performance gap, making them increasingly viable for practical phishing detection scenarios. 

\section{Discussion}
\label{sec:CBA}
The goal of this study was to evaluate how small, non-fine-tuned \Glspl{LLM} perform on a phishing detection task.
While the phishing detection performance of \Glspl{SLM} is promising, these models still lag behind proprietary large-scale models. 
In the following, we discuss the advantages and disadvantages of both approaches, highlighting the trade-offs related to performance, feasibility, costs, and data privacy.

\subsection{Costs} 
One of the primary advantages of small local \Glspl{LLM} is their low operational cost. Running models locally eliminates recurring API fees that are typical of proprietary services, making them particularly attractive for organisations with limited budgets that aim for large-scale deployments, where models are queried almost continuously. The following hosting options are typically available:

\begin{itemize}
    \item \textbf{Proprietary API}: In related literature, GPT4o or GPT4-turbo are frequently used for phishing detection.
    At the time of writing, GPT4o is priced at \$2.5 per million input tokens and \$10 per million output tokens, while GPT4-turbo costs \$10 per million input tokens and \$30 per million output tokens. 
    The newly released GPT5 models are priced at a similar level~\cite{openaipricing}.
    
    Across both experiments in this study, each model processed approximately 8 million input tokens. While output token counts varied between models, using gpt-oss as an example, which belongs to the same family as GPT4, generated roughly 1.6 million output tokens in total.
    Using these figures, the total cost of running the experiments would be approximately \$36 (31€) for GPT4o, or \$128 (109€) for GPT4o-turbo.

    \item \textbf{Renting GPUs}: In this study, models were executed on rented GPUs via runpod, with the following rates: Nvidia A100, 1.4€ per GPU hour, and Nvidia H100, 2.03€ per GPU hour. The total runtime for Experiment 1 and 2 was slightly over 28 hours, resulting in a total cost of approximately 41€ for the A100 setup, including disk storage fees. For the two largest 70b models, which required H100 GPU, the total runtime was 22 hours and 20 minutes, amounting to roughly 46€, including disk storage.

    \item \textbf{Fully local Setup}: As the Nvidia A100 is no longer widely available, the latest GPU generation (H100 series) is an option, which costs between 27,000€ to 39,000€, depending on the provider and specifications ~\footnote{\url{https://www.newegg.com/p/pl?d=h100} (11 September 2025)},\footnote{\url{https://geizhals.at/nvidia-h100-nvl-900-21010-0020-000-a3356480.html} (11 September 2025)}. 
    In addition to the initial investment, electricity, cooling, and maintenance costs must be factored in, which can vary significantly by region and infrastructure.
\end{itemize}

    Taking all these factors into account, the break-even point where investing in local hardware becomes more economical than relying on proprietary APIs ranges between 625 million and 3.6 billion tokens (input and output combined). This range depends on both the GPU price and the selected GPT model. With a high GPU price and a cheaper GPT version, the hardware costs only balance out after a substantial number of tokens are processed. 
    For perspective, in this study, gpt-oss processed about 9.6 million tokens to analyse 2,400 websites. At this rate, the break-even point would be reached after roughly 160,000 to 900,000 analysis runs.
    Considering that the APWG reported over 1 million phishing websites in Q2 2025 alone~\cite{intro1}, running a GPU with local \Glspl{LLM} becomes economically viable compared to using commercial models within just a few months.
    However, the results in \Cref{sec:ex2res} show that models take several seconds to analyse a website. Therefore, (repeated) scans of hundreds of thousands of websites call for more efficient models or, more likely, parallelised analysis setups. Considering the costs for the three approaches, none of them remains economically feasible.
    A potential solution is to reduce the number of websites to analyse by combining classical \Gls{ML} approaches as preselectors and LLMs as final decision-makers.

\subsection{Benefits of SLMs}
The following advantages make (locally-hosted) \Glspl{SLM} attractive for phishing detection and similar security-critical applications.

\begin{itemize}
    \item \textbf{Data Privacy, Control and Security:} Running all inference on local infrastructure ensures that sensitive information such as URLs, HTML content, and user metadata remains internal and is not transmitted to external providers. This is particularly important for organisations subject to strict data protection regulations or operating in sensitive domains.
    In the context of phishing detection, local models allow organisations to maintain full control over their data, reducing exposure to external systems.
    Furthermore, keeping the analysis methods and intellectual property (IP) in-house provides a competitive advantage, preventing sensitive prompts and detection strategies from being shared with third parties. It also lowers the risk of manipulation or infiltration by adversaries.
    
    \item \textbf{Customisability:} Although models in this study were evaluated in their out-of-the-box state, local models can be fine-tuned for phishing detection. Organisations with relevant expertise can leverage their proprietary data to partially retrain model weights or implement a retrieval-augmented generation (RAG) system to improve performance.
    The current open-source \Gls{LLM} ecosystem offers a wide variety of models for domain-specific adaptations. Furthermore, fine-tuned models for related tasks are often shared on platforms such as Hugging Face.
    However, to the best of our knowledge, no fine-tuned models specifically for phishing detection are publicly available at this time.
    
    \item \textbf{Independence and Availability:} Local model deployment eliminates vendor lock-in, ensuring that operations are not dependent on the availability, pricing policies, or strategic decisions of external providers. 
    Local models can also offer greater reliability, as they are not affected by cloud service outages or external network issues. In addition, they can deliver lower latency and faster response times, which is particularly beneficial for time-critical applications such as real-time phishing detection.
\end{itemize}

\subsection{Challenges of SLMs}
Despite these benefits, small local \Glspl{LLM} come with a set of limitations that must be acknowledged.

\begin{itemize}
    \item \textbf{Performance:} The most evident drawback is the lower performance of small models compared to larger proprietary ones. The results achieved in this study are promising; the tested 70b models, for instance, demonstrated solid overall performance, while other models excelled in specific areas such as precision or recall.  
    However, the best proprietary large models used in related work outperformed local models across all performance metrics. 
    Although differences in datasets and detection pipelines make direct comparisons difficult, it is reasonable to assume that a similar performance gap would persist in this experimental setup.
    This gap has practical implications, leading to higher rates of false positives or false negatives. These errors can have severe consequences in security-critical applications such as phishing detection, including missed threats or unnecessary disruptions.

    \item \textbf{Customisability:} Customisability is one of the key strengths of local models, but realising it is non-trivial. First, a suitable base model must be selected, which requires in-depth knowledge of available models and careful evaluation of their strengths and weaknesses. Second, high-quality, domain-specific data is needed to achieve a meaningful fine-tuning result. Finally, depending on the specific approach, fine-tuning itself can be a computationally intensive process, requiring specialised hardware and expertise.

    \item \textbf{Hardware cost:} Running even relatively small models locally requires specialised hardware, such as GPUs or high-performance CPUs, and the technical expertise to set up, optimise, and maintain the infrastructure. For organisations without existing machine learning infrastructure, the initial hardware investment can be substantial. In short-term or exploratory projects, this upfront investment may be too high, making proprietary models or rental GPU services more cost-effective alternatives. Ultimately, the exact cost must be carefully calculated based on the organisation's specific strategy.

    \item \textbf{Scalability:} Proprietary cloud-based solutions offer elastic scaling with demand. In contrast, local deployments are constrained by on-site hardware capacity. Handling sudden spikes in phishing detection workloads may require over-provisioning hardware or accepting degraded performance.
\end{itemize}

\subsection{Application Recommendation}
Based on the highlighted advantages and disadvantages in {\Cref{sec:CBA}}, determining a single optimal setup is challenging, as the decision depends heavily on organisational priorities and operational constraints. With the results of the current study, two primary paths emerge:

\paragraph{Proprietary models for high performance}
If high performance and accuracy are the primary objectives and economic considerations are less critical, proprietary models are the preferred choice. This is particularly true when the dataset size of websites to analyse is small to moderate, or if the analysis is only performed infrequently. Ease of deployment, scalability, and superior performance outweigh the recurring costs.
    
\paragraph{Local \Glspl{LLM} for cost efficiency and privacy} 
The experiments conducted in this work demonstrate that there is potential in running smaller local \Glspl{LLM} for phishing detection, particularly when cost, data control, and privacy are key priorities. Using local models in combination with customisation techniques could become sustainable long-term solutions.
While the initial investment for local model deployment may be substantial, these upfront costs can be offset over time.
Furthermore, having full ownership of the model enables continuous improvements through iterative fine-tuning.

Among the tested local models, llama3.3:70b, deepseek-r1:70b and mistral-nemo:12b stand out as the most promising options out of the box.
The two 70b models demonstrated particularly strong performance when given raw HTML from websites and their corresponding URLs. While all three models are small enough to be fine-tuned at reasonable computational cost, especially the Mistral model, which only consists of 12 billion parameters, is a promising candidate for testing fine-tuned setups. Regarding analysis runtime, the Llama and Gemma models are very similar, while the Mistral model, among the fastest three tested models, delivers its analysis rapidly.

Another consideration that needs to be accounted for is the type of response variable that the models should use. In Experiment 2, we observed that relying on a Boolean classification decision yielded better results than relying on the numeric score. A potential reason may originate from the chosen few-shot prompt design. Models may base their score just on the provided examples by matching them to the techniques found in the HTML code. Due to the nature of few-shot prompting, the list of provided examples is not exhaustive, which means models may miss phishing techniques not included and, therefore, receive a worse score.
The llama3.2:1b model at some point responded with a Python script, which basically counted the number of occurrences of the phishing techniques mentioned in the prompt.

Generally, models with high runtime variability are harder to manage in production environments, as their runtime behaviour is less consistent and may lead to uncertainty in planning.

\section{Limitations}
Although we carefully conducted this work, we acknowledge some limitations.

A more detailed investigation into the linguistic patterns of the reasoning outputs could provide additional insights. Potentially, more advanced \Gls{NLP} methods or even \Glspl{LLM} could be used to analyse the text more thoroughly. However, such analysis is beyond the scope of this work.
In this study, we focused exclusively on HTML code as input for phishing detection models. This choice enabled the inclusion of a broader range of \Glspl{SLM} and allowed for clearer, more controlled comparisons between models.
While we expect performance improvements from incorporating multimodal data (e.g., text and screenshots), a careful analysis of such approaches is left for future research.

\section{Conclusion}
In this work, we evaluated and benchmarked 15 small, local (non-fine-tuned) \Glspl{LLM} for the task of website phishing detection. 
The tested models represent a diverse selection of frequently used architectures, ranging 
from 1b to 70b parameters. Over the course of two experiments, we systematically assessed analysis runtime, output coherence and classification performance. Of the 15 models, nine were deemed practical and provided accurate results on a small sample dataset. These models were then further evaluated on a larger dataset of 1,000 websites.

The results show that overall classification accuracy for most models ranges from 56\% to 89\%, with most models achieving 80\% or more.
However, more substantial differences were observed in precision. For example, the gpt:oss model achieved an impressive precision of 98\% in test runs, but numerous missing JSON outputs and a low recall score hinder its applicability; it was therefore excluded. Models with a more balanced F1-score, such as llama3.3:70, deepseek-r1:70b or mistral-nemo:12b, are more practical in most use cases, as they strike a balance between relatively high accuracy and moderate error rates.

A key observation is that even some mid-sized models in the 10b-20b parameter range achieve F1 Scores comparable to those of previous generations of 70b models. While the current-generation 70b models can outperform proprietary models such as GPT3.5. This reflects a promising trend toward smaller, more efficient models becoming increasingly viable for phishing detection.
Tracking future developments of these models will be particularly interesting.

While the out-of-the-box local models tested in this work do not fully match the performance of state-of-the-art large proprietary \Glspl{LLM}, they offer several advantages, including greater data privacy, customisation potential, and operational independence. Regarding deployment cost, \Glspl{SLM} become a viable option when a heavy workload over an extended period of time is expected, as then the upfront cost of the hardware level out with the API cost of proprietary alternatives. These benefits underscore the importance of conducting further research to enhance classification performance. 

\section{Future Work}
In this work, we provide runtime and performance benchmarks of out-of-the-box (local) \Glspl{SLM} for phishing detection. While the results offer valuable insights, several opportunities for future research remain. First, customisation is a key advantage of open-source models. Future work could focus on fine-tuning these models or integrating RAG techniques to improve performance. 
Second, the dataset used in this work includes additional information, such as website screenshots and other website components, which could be utilised for (multimodal) phishing detection. 
Finally, integrating \Glspl{LLM} into a larger framework for phishing detection, combining multiple tools and analysis stages, could further improve the results. Studying efficient architectures and workflows is subject to future work.

\section*{Acknowledgments}
This work was supported by the Austrian Research Promotion Agency (FFG) under the ``Digitale Technologien 2023'' programme (FFG Project No. FO999915293 -- LLM4CTI).
SBA Research (SBA-K1 NGC) is a COMET Center within the COMET – Competence Centers for Excellent Technologies Programme and funded by BMIMI, BMWET, and the federal state of Vienna. The COMET Programme is managed by FFG.
We also thank Johannes Kosche for working on the infrastructure and assisting with early experiments.
\newpage
\onecolumn
\newpage
\appendix
\section{}
\label{appendix}

Prompt Template 1 shows the complete, non-truncated prompt template used in this work, which includes the detailed list of the few-shot examples provided to the model.
\begin{promptbox}\label{pr:two}
    You are a cybersecurity expert analysing websites for phishing attempts. Your task is to examine the provided HTML code and the original website URL to determine if the website is likely a phishing site.\\
\newline
    **Important:** The HTML may be truncated to reduce costs, so CSS styles and JavaScript code may be missing. Focus on the HTML structure, text content, URLs, and the original URL.\\
    Original URL: '\{original\_url\}'\\
    HTML: '\{html\_text\}'\\
    Original HTML character count: '\{original\_character\_count\}'\\
\newline
    When assessing the likelihood of phishing, consider both your general knowledge of phishing techniques and the evidence you see in the data - focus on HTML structure, content, and domain/URL consistency.\\" 
    Below are common examples of phishing indicators you may find useful, but you are not limited to them:\\
    \newline
    1. **Suspicious URLs/domains** - Check href attributes, form actions, image sources for:\\
        - Misspelled brand names, unusual domains, suspicious subdomains\\
        - IP addresses instead of domains, excessive hyphens, unusual TLDs\\
        
    2. **Form analysis** - Login/input forms with:\\
        - Action URLs pointing to wrong domains\\
        - Password/sensitive data collection for mismatched brands\\
        - Excessive personal information requests (SSN, full address, etc.)\\
        
    3. **Content and language** - Text containing:\\
        - Urgent threats: 'Account suspended', 'Verify immediately', 'Limited time'\\
        - Fear tactics: 'Security breach', 'Unauthorised access detected'\\
        - Reward baits: 'You have won', 'Free gift', 'Exclusive offer'\\
        
    4. **HTML structure issues**:\\
        - Spelling/grammar errors in text content\\
        - Inconsistent or poor HTML structure\\
        - Missing or suspicious meta tags (title, description)\\
        
    5. **Link analysis** - Check all href attributes for:\\
        - Links to different domains than expected\\
        - Shortened URLs (bit.ly, tinyurl, etc.)\\
        - Misleading anchor text vs actual URL\\
        
    6. **Brand impersonation** - Look for:\\
        - Company names in text that don't match domain\\
        - References to legitimate services (PayPal, Amazon, banks, ...) on wrong domains\\
        - Official-sounding but incorrect terminology\\
        
    7. **Missing legitimacy markers**:\\
        - No contact information or privacy policy links\\
        - Missing proper company details in footer\\
        - No legitimate copyright notices\\
    \newline    
    **Note:** Since CSS/JS may be truncated, focus on HTML tags, text content, and URL analysis rather than visual styling or dynamic behaviour.\\
    **Scoring guide:**\\
    - 0-2: Very unlikely phishing (legitimate site)\\
    - 3-4: Low risk (minor suspicious elements)\\
    - 5-6: Medium risk (several concerning indicators)\\
    - 7-8: High risk (multiple clear phishing signs)\\
    - 9-10: Very high risk (obvious phishing attempt)\\
   \newline     
    **Required output format (JSON only):**\\
    \{\\
        phishing\_score: int [0-10],\\
        is\_phishing: boolean [true/false],\\
        reasoning: string [Brief explanation of your decision based on specific indicators found]\\
    \}\\
    \newline    
    **Output Constraints:**\\
    Do only output the JSON-formatted output and nothing else.

    \end{promptbox}

\Cref{tab:avgruntime} is referenced in \Cref{sec:runtime}, where only the fastest and slowest models are discussed. 
\begin{table}[ht]
\caption{Average analysis runtime (in seconds) per model and website run for the D5 and the D50 dataset. The table is sorted by dataset and in ascending order according to runtime.}
\label{tab:avgruntime}
\begin{tabular}{lll}
\toprule
\textbf{Dataset} & \textbf{Model} & \textbf{Runtime} \\
\midrule
d5 & llama3.2:1b & 0.550 \\
d5 & dolphin3:8b & 0.892 \\
d5 & mistral-nemo:12b & 1.793 \\
d5 & gemma3:4b & 1.966 \\
d5 & deepseek-r1:1.5b & 2.916 \\
d5 & gemma3:12b & 3.545 \\
d5 & phi3:14b & 4.515 \\
d5 & qwen3:4b & 5.725 \\
d5 & llama3.1:8b & 5.729 \\
d5 & gpt-oss:20b & 5.784 \\
d5 & deepseek-r1:14b & 6.386 \\
d5 & gemma3:27b & 7.077 \\
d5 & llama3.3:70b & 7.731 \\
d5 & qwen3:30b & 9.375 \\
d5 & deepseek-r1:70b & 23.318 \\
d50 & llama3.2:1b & 0.876 \\
d50 & dolphin3:8b & 2.162 \\
d50 & mistral-nemo:12b & 3.358 \\
d50 & deepseek-r1:1.5b & 3.752 \\
d50 & gemma3:4b & 3.840 \\
d50 & deepseek-r1:14b & 4.618 \\
d50 & llama3.1:8b & 5.353 \\
d50 & gemma3:12b & 7.846 \\
d50 & gpt-oss:20b & 7.946 \\
d50 & qwen3:4b & 8.601 \\
d50 & phi3:14b & 8.913 \\
d50 & qwen3:30b & 12.709 \\
d50 & gemma3:27b & 15.658 \\
d50 & llama3.3:70b & 15.748 \\
d50 & deepseek-r1:70b & 30.775 
\end{tabular}
\end{table}

The qualitative reasoning output analysis in \Cref{tab:quali_ana} is part of \Cref{sec:outputstab} and contains the complete information of the phishing and benign wording tendencies per model and dataset.
\begin{table}[ht]
\caption{Overview of the qualitative output analysis per model and dataset, summarised in the Phishing Score and Benign Score tendencies.}
\label{tab:quali_ana}
\begin{tabular}{llrr}
\toprule
\textbf{Model} & \textbf{Dataset} & \textbf{Phishing Score Tendency} & \textbf{Benign Score Tendency} \\
\midrule
deepseek-r1:1.5b & d5 & 336 & 86 \\
deepseek-r1:14b & d5 & 322 & 109 \\
deepseek-r1:70b & d5 & 333 & 127 \\
dolphin3:8b & d5 & 257 & 125 \\
gemma3:12b & d5 & 485 & 212 \\
gemma3:27b & d5 & 483 & 186 \\
gemma3:4b & d5 & 518 & 123 \\
gpt-oss:20b & d5 & 194 & 95 \\
llama3.1:8b & d5 & 116 & 19 \\
llama3.2:1b & d5 & 10 & 0 \\
llama3.3:70b & d5 & 306 & 151 \\
mistral-nemo:latest & d5 & 219 & 166 \\
phi3:medium & d5 & 304 & 117 \\
qwen3:30b & d5 & 14 & 2 \\
qwen3:4b & d5 & 101 & 106 \\
deepseek-r1:1.5b & d50 & 207 & 31 \\
deepseek-r1:14b & d50 & 301 & 166 \\
deepseek-r1:70b & d50 & 318 & 176 \\
dolphin3:8b & d50 & 197 & 83 \\
gemma3:12b & d50 & 475 & 190 \\
gemma3:27b & d50 & 425 & 195 \\
gemma3:4b & d50 & 470 & 115 \\
gpt-oss:20b & d50 & 107 & 63 \\
llama3.1:8b & d50 & 170 & 19 \\
llama3.2:1b & d50 & 18 & 0 \\
llama3.3:70b & d50 & 246 & 113 \\
mistral-nemo:latest & d50 & 156 & 112 \\
phi3:medium & d50 & 235 & 57 \\
qwen3:30b & d50 & 0 & 0 \\
qwen3:4b & d50 & 164 & 239 
\end{tabular}
\end{table}
\pagebreak

Model Output 1 shows the typical result of the DeepSeek models, highlighting the combination of the thinking process and the required JSON output. Although other models used in this work are also reasoning models, only the DeepSeek models provide a thinking process in this way. 
\begin{resultbox}[model:M1]
    "\}\textbackslash n\textbackslash nOkay, so I need to analyze this website's HTML code for phishing attempts. The user provided an HTML snippet that seems very minimal—just a title link. But since CSS and JS might be missing, I have to focus on the structure and content.\textbackslash n\textbackslash nFirst, let me look at the HTML structure. It starts with 
    $<$html$>$, then $<$head$>$ which includes meta tags, charset, and a title. The body is empty except for some empty $<$body$><$/body$><$/html$>$. So, no actual content in the page beyond the title link.\textbackslash n\textbackslash nNow, looking at the phishing indicators:\textbackslash n\textbackslash n1. **Suspicious URLs/Domain**: The HTML doesn't have any links or href attributes because it's just an empty body. But wait, maybe there are other elements? No, the structure is very minimal—probably no form fields or images either since they're not present.\textbackslash n\textbackslash n2. **Form Analysis**: There are no login inputs or forms in this snippet. So, that's probably not a concern here.\textbackslash n\textbackslash n3. **Content and Language**: The only text is the title: \textbackslash"Amazon\textbackslash". That doesn't seem like it contains any phishing indicators. It's just a link to an account sign-in page for Amazon. I don't see any urgency, threats, or reward baits here.\textbackslash n\textbackslash n4. **HTML Structure Issues**: There are no errors in the HTML structure because there are no tags beyond what's already present. So, that doesn't add anything.\textbackslash n\textbackslash n5. **Link Analysis**: The only link is to \textbackslash"Amazon\textbackslash", which is a legitimate account sign-in page for Amazon. It doesn't point to any other domain or have any suspicious links.\textbackslash n\textbackslash n6. **Brand impersonation**: The text has \textbackslash"Amazon\textbackslash" in the title, but it's just a regular brand name. There are no references to other companies or services on this page. So, that's not a phishing indicator either.\textbackslash n\textbackslash n7. **Missing Legitimacy Markers**: The HTML doesn't have any meta tags like title or description. It does have a link tag, which is fine because it's pointing to a legitimate domain. There are no proper company details in the footer or any privacy policy links. So, that's not an issue here.\textbackslash n\textbackslash nPutting this all together: The website has very minimal HTML structure and content. It doesn't contain any phishing indicators like suspicious URLs, form fields, incorrect domains, or references to other companies. The only link is to a legitimate account sign-in page for Amazon, which isn't phishing material.\textbackslash n\textbackslash nTherefore, based on the analysis, there are no clear signs of phishing attempts here. So, the phishing score should be 0-2, indicating very unlikely.\textbackslash n</think>\textbackslash n\textbackslash n```json\textbackslash n\{\textbackslash n  \textbackslash"phishing\_score\textbackslash": 0,\textbackslash n  \textbackslash"is\_phishing\textbackslash": false,\textbackslash n  \textbackslash"reasoning\textbackslash": \textbackslash"The website's HTML code is minimal and lacks any links or content that would indicate phishing attempts. The only link points to a legitimate Amazon account sign-in page, which does not contain phishing indicators.\textbackslash"\textbackslash n\}\textbackslash n```"
\end{resultbox}

Model Output 2 illustrates a common occurrence, where the model's first output differs from all subsequent runs, which are then identical. 
\begin{resultbox}[model:M2]

    \textbf{Run 1}: '```json\textbackslash n\{\textbackslash n"phishing\_score": 3,\textbackslash n"is\_phishing": false,\textbackslash n"reasoning": "The HTML code is minimal and lacks significant content. It contains basic HTML tags for a webpage, but there are no suspicious URLs, forms, or text content that would immediately indicate a phishing attempt. The lack of content makes it difficult to assess the legitimacy of the site."\textbackslash n\}\textbackslash n```'\\
    \newline
   \textbf{Run 2}: "```json\textbackslash n\{\textbackslash n\textbackslash"phishing\_score\textbackslash": 3,\textbackslash n\textbackslash"is\_phishing\textbackslash": false,\textbackslash n\textbackslash"reasoning\textbackslash": \textbackslash"The HTML code is minimal and lacks significant content. It contains basic HTML tags for a webpage, but no suspicious URLs, forms, or text content are present. It's difficult to assess the legitimacy of the site without more information. The score reflects the lack of indicators, but it's not a definitive determination of legitimacy.\textbackslash"\textbackslash n\}\textbackslash n```"'\\
    \newline
    \textbf{Run 3}: "```json\textbackslash n\{\textbackslash n\textbackslash"phishing\_score\textbackslash": 3,\textbackslash n\textbackslash"is\_phishing\textbackslash": false,\textbackslash n\textbackslash"reasoning\textbackslash": \textbackslash"The HTML code is minimal and lacks significant content. It contains basic HTML tags for a webpage, but no suspicious URLs, forms, or text content are present. It's difficult to assess the legitimacy of the site without more information. The score reflects the lack of indicators, but it's not a definitive determination of legitimacy.\textbackslash"\textbackslash n\}\textbackslash n```"'\\
    \newline
    \textbf{Run 4}: "```json\textbackslash n\{\textbackslash n\textbackslash"phishing\_score\textbackslash": 3,\textbackslash n\textbackslash"is\_phishing\textbackslash": false,\textbackslash n\textbackslash"reasoning\textbackslash": \textbackslash"The HTML code is minimal and lacks significant content. It contains basic HTML tags for a webpage, but no suspicious URLs, forms, or text content are present. It's difficult to assess the legitimacy of the site without more information. The score reflects the lack of indicators, but it's not a definitive determination of legitimacy.\textbackslash"\textbackslash n\}\textbackslash n```"'\\
    \newline
    \textbf{Run 5}: "```json\textbackslash n\{\textbackslash n\textbackslash"phishing\_score\textbackslash": 3,\textbackslash n\textbackslash"is\_phishing\textbackslash": false,\textbackslash n\textbackslash"reasoning\textbackslash": \textbackslash"The HTML code is minimal and lacks significant content. It contains basic HTML tags for a webpage, but no suspicious URLs, forms, or text content are present. It's difficult to assess the legitimacy of the site without more information. The score reflects the lack of indicators, but it's not a definitive determination of legitimacy.\textbackslash"\textbackslash n\}\textbackslash n```"'\\
\end{resultbox}

\printglossary
\clearpage

\end{document}